\documentclass[journal]{IEEEtran}
\usepackage{graphicx}
\usepackage{amstext}
\usepackage{algorithm}
\usepackage{algorithmic}
\usepackage{mathrsfs}
\usepackage{amssymb}
\usepackage{amsmath}
\usepackage{bm}
\allowdisplaybreaks[4]
\usepackage{epstopdf}
\usepackage{multicol}
\usepackage{stfloats}
\usepackage{url}
\usepackage{color}
\usepackage{enumerate}
\usepackage{breqn}
\usepackage{bbm}
\usepackage{cite}
\usepackage{caption}
\usepackage{enumitem}
\usepackage{array}
\ifCLASSINFOpdf
\else
\fi
\begin{document}

\title{Blockchain for Future Smart Grid: A Comprehensive Survey}

\author{Muhammad Baqer Mollah,~\IEEEmembership{Member,~IEEE,}
        Jun Zhao,~\IEEEmembership{Member,~IEEE,}
		Dusit Niyato,~\IEEEmembership{Fellow,~IEEE,}
        Kwok-Yan Lam,~\IEEEmembership{Senior Member,~IEEE,}
        Xin Zhang,~\IEEEmembership{Member,~IEEE,}
        Amer M.Y.M. Ghias,~\IEEEmembership{Member,~IEEE,}
        \\Leong Hai Koh,~\IEEEmembership{Member,~IEEE,}
        and Lei Yang,~\IEEEmembership{Senior Member,~IEEE}%
\thanks{Citation information: DOI: https://doi.org/10.1109/JIOT.2020.2993601}
\thanks{The work of J. Zhao is supported by the Nanyang Technological University (NTU) Startup Grant M4082311.020, Alibaba-NTU Singapore Joint Research Institute (JRI) M4062640.J4I, Singapore Ministry of Education Academic Research Fund Tier 1 RG128/18, RG115/19, and Tier 2 MOE2019-T2-1-176, and NTU-WASP Joint Project M4082443.020. The work of D. Niyato is supported by the National Research Foundation (NRF), Singapore, under Singapore Energy Market Authority (EMA), Energy Resilience, NRF2017EWT-EP003-041, Singapore NRF2015-NRF-ISF001-2277, Singapore NRF National Satellite of Excellence, Design Science and Technology for Secure Critical Infrastructure NSoE DeST-SCI2019-0007, A*STAR-NTU-SUTD Joint Research Grant on Artificial Intelligence for the Future of Manufacturing RGANS1906, WASP/NTU M4082187 (4080), Singapore MOE Tier 2 MOE2014-T2-2-015 ARC4/15, and MOE Tier 1 2017-T1-002-007 RG122/17. The work of K. Y. Lam is supported by the National Research Foundation, Prime Minister’s Office, Singapore under its Strategic Capability Research Centres Funding Initiative. The work of L. Yang is supported in part by the U.S. National Science Foundation under Grants EEC-1801727, IIS-1838024, and CNS-1950485. \textit{(Corresponding author: Muhammad Baqer Mollah)}}
\thanks{M. B. Mollah, J. Zhao, D. Niyato, and K. Y. Lam are with the School of Computer Science and Engineering, Nanyang Technological University, Singapore 639798 (Email: muhd.baqer@ntu.edu.sg; junzhao@ntu.edu.sg; dniyato@ntu.edu.sg; kwokyan.lam@ntu.edu.sg).}%
\thanks{X. Zhang and A. M. Y. M. Ghias are with the School of Electrical and Electronic Engineering, Nanyang Technological University, Singapore 639798 (Email: jackzhang@ntu.edu.sg; amer.ghias@ntu.edu.sg).}%
\thanks{L. H. Koh is with the Energy Research Institute, Nanyang Technological University, Singapore 639798 (Email: lhkoh@ntu.edu.sg).}%
\thanks{L. Yang is with the Department of Computer Science and Engineering, University of Nevada, Reno, NV 89557, USA (Email: leiy@unr.edu).}%
\thanks{Copyright (c) 20xx IEEE. Personal use of this material is permitted. However, permission to use this material for any other purposes must be obtained from the IEEE by sending a request to pubs-permissions@ieee.org.}
}

\markboth{IEEE Internet of Things Journal 2020}%
{Shell \MakeLowercase{\textit{et al.}}: Bare Demo of IEEEtran.cls for IEEE Journals}

\maketitle

\begin{abstract}
The concept of smart grid has been introduced as a new vision of the conventional power grid to figure out an efficient way of integrating green and renewable energy technologies. In this way, Internet-connected smart grid, also called energy Internet, is also emerging as an innovative approach to ensure the energy from anywhere at any time. The ultimate goal of these developments is to build a sustainable society. However, integrating and coordinating a large number of growing connections can be a challenging issue for the traditional centralized grid system. Consequently, the smart grid is undergoing a transformation to the decentralized topology from its centralized form. On the other hand, blockchain has some excellent features which make it a promising application for smart grid paradigm. In this paper, we aim to provide a comprehensive survey on application of blockchain in smart grid. As such, we identify the significant security challenges of smart grid scenarios that can be addressed by blockchain. Then, we present a number of blockchain-based recent research works presented in different literatures addressing security issues in the area of smart grid. We also summarize several related practical projects, trials, and products that have been emerged recently. Finally, we discuss essential research challenges and future directions of applying blockchain to smart grid security issues.
\end{abstract}


\IEEEpeerreviewmaketitle

\section{Introduction}
	\IEEEPARstart{I}{n} the past few decades, traditional centralized fossil fuel-based energy systems have been facing some major challenges such as long-distance transmission, carbon emission, environment pollution, and energy crisis. In order to build a sustainable society by addressing these challenges, utilization of renewable energy from diverse sources as well as improving the efficiency of energy usage are the two key potential solutions. In recent years, the smart grid concept \cite{dileep2020survey, kakran2018smart, rehmani2018integrating, dragivcevic2019future, shaukat2018survey, shaukat2018survey2} which involves communication technology, interconnected power system, advanced control technology, and smart metering has been applied to improve the utilization of renewable energy sources and relieve the energy crisis somehow. The concept of smart grid has been introduced as a new vision of conventional power grid which offers two-way energy and information exchange in order to figure out an efficient way of delivering, managing, and integrating green and renewable energy technologies.
	
	Unfortunately, the smart grid makes it difficult to enhance the access to distributed and scalable energy resources at a large scale as well as ensure energy security and integrate other approaches to improve the energy utilization efficiency and reliability. Therefore, in order to advance it and solve the current limitations, the Energy Internet (EI), also called Internet of Energy (IoE) or Smart Grid 2.0, has been introduced by integrating smart grid context with Internet technology \cite{mahmud2020internet, al2019iot, saleem2019internet, kabalci2019internet, wang2017survey, yijia2018comprehensive, dong2014smart, chen2019internet, hussain2019emerging}. In contrast with the smart grid, the EI is an Internet-style solution for energy related issues by accommodating with IoT, advanced information \& communication technologies, power system components, and other energy networks. The aim of this emerging and innovative approach is to ensure the connection of energy anywhere at any time. In summary, both concepts have been developed with aims to ensure that all the participants and components have the ability (i) to interact closely with each other, (ii) to make decisions by themselves, (iii) to exchange both energy and associated information in multiple ways, (iv) to access large-scale different types of distributed energy resources seamlessly, (v) to adapt with both centralized and distributed energy sources, (vi) to balance energy supply and demand through energy  sharing, and (vii) to ensure flexible energy generation/selling and purchasing/consuming the energy. Since the connectivity is becoming larger, a major challenging issue is to integrate and coordinate a large number of connections such as growing distributed energy producers, their consumers, electric vehicles, smart devices, and cyber-physical system within the traditional centralized grid system. Managing such continuously growing network in a centralized manner will require sophisticated and costly information \& communication infrastructures. Thus, moving towards decentralization is a trend in smart grid so that all its components can incorporate and integrate in a dynamic way. Also, decentralization is one of the fundamental requirements in the EI’s development happening in smart grid according to its vision.

	However, the decentralized smart grid system with a large number of components and complex connections may also be a security, privacy, and trust nightmare which requires new and innovative technologies to address \cite{gunduz2020cyber, islam2019physical, kumar2019smart, ghosal2019key, de2019security, rastogi2019toward, shayeghi2019survey}. On the other hand, as an emerging and promising technology, blockchain offers new opportunities to make decentralized systems. This blockchain technology is decentralized, i.e., to manage blockchain, no central trusted authority is required; instead, multiple entities in the network can do among themselves to create, maintain, and store a chain of blocks. Every entity can verify that the chain order and data have not been tampered with. This decentralized system makes any system redundant and resilient to system failure \& cyber-attacks and solve many problems of centralized system. Although the blockchain is initially introduced and populated as digital currencies \cite{nakamoto2008bitcoin, wood2014ethereum}, due to having its excellent properties, it is attracting enormous attention in many other non-monetary applications. At the same time, beyond digital currencies, blockchain is also promoting the realization of secure, privacy-preserving, and trusted smart grid developments toward decentralization.

	\textit{Related Surveys and Our Contributions:} Though blockchain in smart grid is a new area of research, it has attracted already significant attention. Recently, several research works have been conducted to address smart grid security, privacy and trust issues by blockchain. Till date, a number of surveys have been published in \cite{wu2018application, energyblockchain, musleh2019blockchain, andoni2019blockchain, wang2019energy, hassan2019blockchain, goranovic2017blockchain, siano2019survey, troncia2019distributed, ahl2019review}, where it has been attempted to review these research works from different angles and scopes. For instance, recent surveys discuss the application of blockchain for Energy Internet (EI)/Internet of Energy (IoE) in \cite{wu2018application, energyblockchain}. Surveys of blockchain on microgrid and overview of related projects are given in \cite{goranovic2017blockchain}. Surveys in blockchain-based peer-to-peer (P2P) energy trading and decentralized energy market are presented in \cite{wang2019energy, siano2019survey, troncia2019distributed, ahl2019review}. A survey on blockchain applications in smart grid along with a new framework is presented in \cite{musleh2019blockchain}, but without discussing any future research opportunities. The work in \cite{andoni2019blockchain} outlines the blockchain potentials and notable use cases in energy applications such as energy trading, microgrids, and electric e-mobility. A survey of potential benefits of blockchain for smart energy system is presented in \cite{hassan2019blockchain}, where related blockchain platforms and projects are discussed. A summary of the contributions of our work with respect to others is presented in Table I. However, these aforesaid works only consider either a particular issue or some selected topics only. There is no survey that covers a broad aspect of smart grid domain in state-of-the-art blockchain research. This motivates us to deliver this paper with the comprehensive and thorough survey on up-to-date activities of rapidly growing blockchain in smart grid research. However, in contrast to the related survey works, the main contributions of this paper are summarized as follows.

\begin{itemize}
  \item We present a brief introduction to blockchain background including definitions of distributed ledger technology, blockchain technology \& smart contract, blockchain categories, and blockchain consensus mechanisms.
  \item We outline major requirements that smart grid must meet. We also summarize the smart grid security, privacy \& trust objectives, and describe how blockchain can address.
  \item We discuss the key research challenges of different parts and scenarios of smart grid domain in order to realize why blockchain should apply, and how blockchain can contribute to solving those challenges.
  \item We discuss the opportunities of blockchain research in the area of smart grid.
  \item We present a comprehensive literature review on various existing blockchain-based solutions developed for smart grid. We also highlight which issues have been investigated, and which techniques have been utilized along with blockchain.
  \item We summarize the recent practical initiative related to blockchain and smart grid.
  \item Based on our study, we point out further challenges that still remain to be addressed and future research directions.
\end{itemize}

\begin{table}[ht]
\centering
\caption{Summary of our contributions compared to other recent works}
\begin{tabular}{m{1.5cm}|m{6.5cm}}

\hline \hline
Work			& Contribution \\
\hline
\cite{wu2018application}		& Blockchain application to Energy Internet

Challenges associated with Blockchain in Energy Internet\\
\hline
\cite{energyblockchain}			& Blockchain in Energy Internet

Initiatives related to blockchain for Energy Internet\\
\hline
\cite{musleh2019blockchain}			& Blockchain applications in smart grid

A proposed theoretical framework

Blockchain and smart grid testbeds\\
\hline
\cite{andoni2019blockchain}			& Blockchain in decentralized energy trading and consumer-centric marketplace

Future energy industry with blockchain

Key questions before adopting blockchain in energy industry\\
\hline
\cite{wang2019energy}		& Energy trading by Blockchain\\
\hline
\cite{hassan2019blockchain}			& Blockchain adoption in smart energy

Blockchain-assisted smart energy projects\\
\hline
\cite{goranovic2017blockchain}	& Blockchain in microgrid network\\
\hline
\cite{siano2019survey}			& Peer-to-Peer (P2P) transactive energy trading\\
\hline
\cite{troncia2019distributed, ahl2019review}			& P2P energy exchange in local distributed network\\
\hline
This Work			& Motivations of adopting blockchain in smart grid

Blockchain for AMI, decentralized energy trading \& market, energy CPS, EVs management, and microgrid

Blockchain in smart grid practical initiatives

Smart grid specific research direction for future works\\
\hline

\end{tabular}
\end{table}	

	\textit{Paper Organization:} The remainder of this paper is outlined as follows. Section II gives an overview of blockchain technology background. Section III considers blockchain in smart grid where we discuss how the smart grid is transforming into the decentralized system, and also, how blockchain can contribute. Section IV presents the recent blockchain contributions in smart grid. Section V summarizes the practical initiatives related to blockchain adoption in smart grid. Section VI discusses important research challenges and future research directions. Finally, section VII concludes the paper. The list of acronyms and their definitions used in this paper is highlighted in the Table II.
	
\begin{table}[ht]
    \centering
    \caption{List of Major Acronyms and their Definitions.}
\begin{tabular}{l|l}
\hline \hline
Acronym                                      & Their Definition \\
\hline
SG                         					 & Smart Grid \\
\hline
EI                      					 & Energy Internet \\
\hline
IoE											 & Internet of Energy \\
\hline
DER											 & Distributed Energy Resources \\
\hline
EV										     & Electric Vehicle \\
\hline
V2G									         & Vehicle to Grid \\
\hline
SCADA									     & Supervisory Control and Data Acquisition \\
\hline
CPS									         & Cyber-Physical System \\
\hline
DSO							                 & Distributed System Operator \\
\hline
ESCO	 									 & Energy Service Company \\
\hline
PMU										     & Phasor Measurement Unit \\
\hline
AMI											 & Advanced Metering Infrastructure \\
\hline
DG									         & Distributed Generator \\
\hline
DLT										     & Distributed Ledger Technology \\
\hline
DAG											 & Directed Acyclic Graph \\
\hline
PoW										     & Proof of Work \\
\hline
PoS											 & Proof of Stake \\
\hline
BFT	 										 & Byzantine Fault Tolerance \\
\hline
P2P											 & Peer to Peer \\
\hline
ECC										     & Elliptic Curve Cryptography \\
\hline
ICS											 & Industrial Control System \\
\hline
ESU	 										 & Energy Storage Unit \\
\hline
ADMM							  		     & Alternating Direction Method of Multipliers \\
\hline
TES									         & Transactive Energy System \\
\hline
CDA									         & Continuous Double Auction \\
\hline
DApp										 & Decentralized Application\\
\hline
\end{tabular}
\end{table}	
	
\section{Blockchain Background}
In this section, we present the blockchain background information for this paper which includes three subsections. These subsections include the definitions of distributed ledger technology, blockchain \& smart contract, blockchain categories, and consensus mechanisms.

	\subsection{DLT, Blockchain, and Smart Contract}
	\textit{DLT:} A distributed ledger technology can be defined as a consensus of replicated, shared, and synchronized digital data. This data is usually spread across numerous sites, countries, and institutions geographically. Also, this data that are shared across the network can be accessed by participant at each node of the network. The changes made to the ledger are brought back and copied to all the participants within seconds or minutes. Moreover, the nodes are capable of updating themselves with the new and corrected copy of the ledger. The cryptographic keys and signatures help to achieve security.
	
	\textit{Blockchain Technology:} The blockchain technology, firstly introduced in \cite{nakamoto2008bitcoin} as a chain of blocks, is a distributed ledger including a collection of blocks that registers different records of data or transaction information. Here, the blocks are attached together with a chain where each block references the cryptographic hash of the previous block’s data. In blockchain network, newly generated blocks are continuously added to the chain at regular intervals, and this chain is replicated among the members of the network. Each block may also include timestamp, nonce, a hash tree named Merkle tree \cite{merkle1987digital}, smart contract scripts \cite{wood2014ethereum}, and so on. The hash and Merkle tree allows verifying that the content inside the block is not modified, i.e., ensuring integrity. In particular, the Markle trees are suitable for lightweight devices that do not have enough space to store the entire blockchain since they allow lightweight devices to search the inclusion of data quickly and verify the data. Moreover, to alter any block’s content, it is required to change all the blocks since the hash of a block becomes different almost surely if any of its content changes, and each block has previous block’s hash which makes it practically impossible to modify the chain maliciously. Apart from this linear chain-based structure, another type of structure named Directed Acyclic Graph (DAG) is also available where each bock references multiple previous blocks.
	
	\textit{Smart Contract:} In the 1990s, Nick Szabo introduced smart contract in \cite{misc1, misc2} as a computerized protocol which is able to execute the terms and conditions of an agreement. Within the blockchain context, a smart contract \cite{buterin2014next} is a computer script which is stored and deployed in blockchain. Instead of legal languages, the smart contract records the conditions and events such as an asset’s targeted value, an ending date, or transaction information. The main feature of this smart contract is that when a condition is met, or an event is reached, the contract can be executed automatically according to the scripts. Since a smart contract is deployed in the blockchain, it can run without any centralized control. The Ethereum \cite{wood2014ethereum} is the most popular smart contract platform based on blockchain. Fig. 1 shows the logical representation of blockchain structures, a typical block structure, and smart contract’s working principle.
	
\begin{figure*}[ht]
    \centering
       \includegraphics[width=1\textwidth]{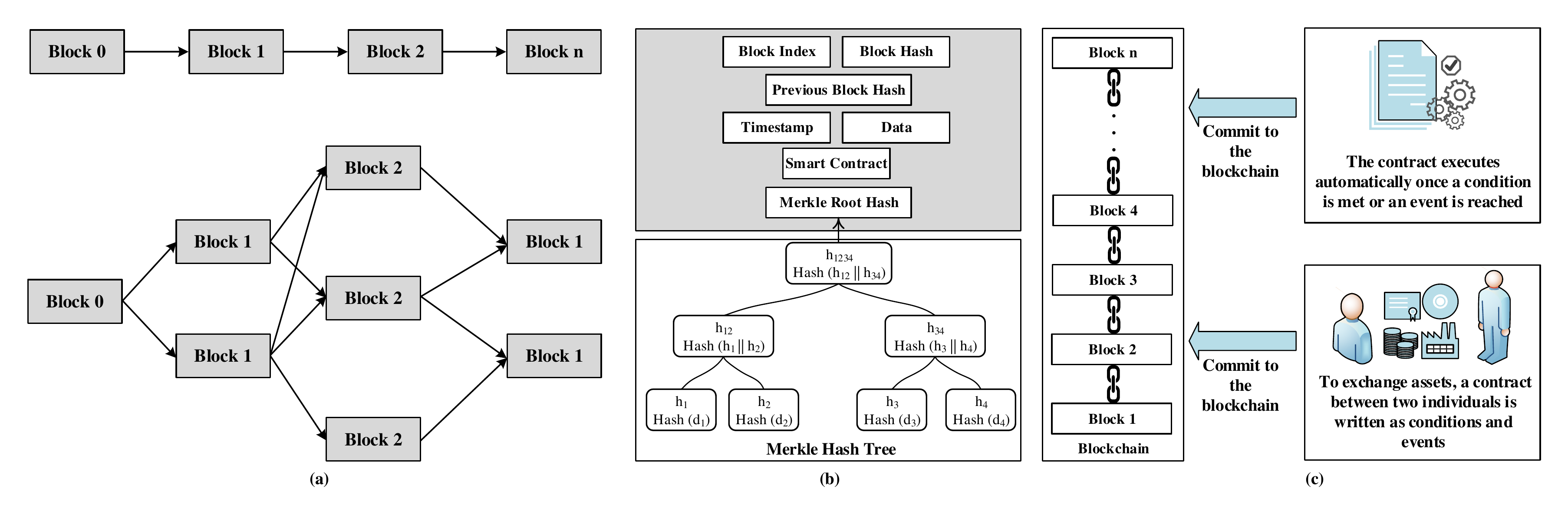}
  \caption{Graphical representation of blockchain structures and working principle of smart contract: (a) Linear-chain based and DAG-based logical structure.  (b) A typical block structure and Merkle hash tree. (c) The basic working principle of smart contract}
\end{figure*}
	
	\subsection{Blockchain Categories}
	\textit{Permissioned vs Permissionless:} Based on how blockchain is restricted to participate in creating new blocks and access the block contents, they can be permissioned or permissionless. In permissionless chain, anyone can join the blockchain network and engage in creating a new block. On the other hand, in permissioned chain, only pre-defined and authorized nodes can do this.
	
	\textit{Private vs. Public:} Blockchain can also be categorized as public and private. The public blockchains are truly decentralized and permissionless. They allow open participation and maintaining a copy of the chain by anyone. Usually, this type of blockchain has a large number of anonymous users. In contrast to the public chains, in the private blockchains, some selected/pre-defined and trusted users are permissioned to validate and participate in publishing the new blocks. Other public or permitted users in the network are restricted to read the data in the blocks. Unlike the public, the private chain may be partially decentralized. Furthermore, another type of private chain is named as consortium or federated blockchain which is also a permissioned chain. In this type of blockchain, a number of organizations make a consortium to maintain the blockchain and allow it to ensure transparency among the participants. Though, the private blockchain is still a centralized network, this kind of blockchain is usually developed to control by an organization and also, to increase the throughput.
	
	\textit{On-chain vs Off-chain:} Blockchain transactions can be on-chain or off-chain. They differ from each other in a number of ways. The transactions available on the blockchain which are visible to all the users on the blockchain network are the on-chain transactions. The transactions are being confirmed by a suitable number of participants. It also involves recording the details of the transaction on an appropriate block and transmission of the basic and essential information to the entire blockchain network.
	
	On the other hand, the off-chain transactions involve the movement of value outside of the blockchain network. They are becoming more popular, particularly among large participants because of their low cost. The first major advantage is that they can be executed instantly. Contrarily, on-chain transactions can have a lengthy lag-time. In the case of on-chain transactions, the lag-time usually depends upon the network load and number of transactions waiting in the chain to be confirmed. Secondly, with off-chain transactions, nothing occurs on the blockchain and no miner or validator is needed to confirm the transaction. Thus, usually, there is no transaction fee. Off-chain transaction is a good choice, particularly when a large number of transactions are involved. The on-chain transactions are often very costly and challenging in micropayment systems where small amounts of payments cannot be transacted due to high transaction fees. Apart from this, in the case of off-chain transactions, details are not publicly broadcasted. Consequently, they offer more security and anonymity to the users. With on-chain transactions, a user’s identity may be possible to derive partially by studying the transaction patterns.	
	
	\subsection{Consensus Mechanisms}
	The consensus mechanisms are one of the key components of blockchain technology in order to add newly published blocks into the blockchain. In both public and private blockchain, a consensus mechanism ensures the trust in the network where a set of validators/miners commonly reaches an agreement whether the block is valid or not. In the public blockchains, anyone can take part in consensus without having the trust of other nodes in the blockchain network. Due to security reasons, the consensus algorithms typically incur very high costs in terms of computational power. On the other hand, due to the trust and a limited number of miners/validators, the consensus mechanisms of private blockchain are relatively simpler than public ones to achieve higher network throughput. In the following, we briefly explain some popular consensus mechanisms below.
	
	\textit{Proof of Work (PoW):} The PoW is the first public blockchain consensus which is introduced in Bitcoin \cite{nakamoto2008bitcoin}. The main idea behind this mechanism is that in order to create a new block, the consensus nodes are asked to solve a computationally expensive puzzle, known as PoW problem which is hard to solve but easy to verify. Once solved, the solution is attached to the new block, and it broadcasts across the network. This attachment allows any other nodes to verify the correctness of a new block published by the particular node. Here, the processing is also called mining, and usually, it is incentive-based. However, though the target of the PoW is to try to avoid different kinds of cyber-attacks, it is also vulnerable to 51\% attack where one or a group of malicious nodes may take control of 51\% of processing power in the blockchain network. Additionally, the mining process of PoW introduces some drawbacks such as inefficient throughput, high latency, and high energy consumption that make PoW unsuitable for many other blockchain applications.
	
	\textit{Proof of Stake (PoS):} The PoS \cite{misc3, nguyen2019proof} is the most popular alternative mechanism of PoW which aims to improve upon PoW’s common limitations. In PoS-based blockchain, the term “mining” is replaced by “validating”, i.e., blocks are commonly validated rather than mined. The concept behind PoS is that the algorithm randomly determines the validators to create the new blocks, and the probability of a node validating the next new block is proportional to the stakes/assets (e.g., coins) it owns. In other words, instead of running high computational puzzle-solving, in PoS, the validators need to prove its share in the network according to the current chain. The PoS is implemented in \cite{king2012ppcoin}. However, the wealthiest validators may be the ones administering the blockchain which makes the PoS mechanism unfair. To overcome this, in \cite{misc4}, the authors consider stake age into consideration. The validators having the oldest and largest assets would be more likely to validate a block.
	
	\textit{Delegated Proof of Stake (DPoS):} The DPoS \cite{misc5} is a variant of PoS, but the main difference between PoS and DPoS is that in DPoS, only a number of selected delegates can generate and validate the blocks. In other words, PoS is directly democratic, whereas, DPoS is delegated democratic. Since fewer nodes can engage in validating process, the DPoS is even faster than PoS. BitShares blockchain \cite{misc6} uses DPoS as part of its consensus mechanism.
	
	\textit{Leased Proof of Stake (LPoS):} The LPoS \cite{misc7} allows the nodes to lease their own assets to others. The main aim of this leasing is to increase the probability of being validators, increase the number of voteable participants and reduce the probability of the blockchain network being ruled by a single group of nodes. Usually, the incentives are shared proportionally.
	
	\textit{Proof of Activity (PoAc):} The PoAc \cite{bentov2014proof} consensus mechanism is developed based on PoW and PoS. The block creators of the next new blocks in the blockchain initially work as miners using PoW mechanism to defense security attacks, and hence, they start to receive the rewards. Once the miners has enough coins (asset), they move to utilize the PoS mechanism to publish new blocks. In \cite{duong20162, chepurnoy2017twinscoin}, almost similar mechanisms are introduced.
	
	\textit{Proof of Burn (PoB):} The PoB \cite{misc8} is an alternative of both PoW and PoS. The PoB allows the validators to create a new block and get rewarded once they burn their own coins/assets by delivering to verifiable, public, and un-spendable addresses. This spending coin is considered an investment. Hence, after investing, a user can make their stakes on the chain and become an authorized validator. In contrast with PoW and PoS, the PoB does not require energy consumption. The Slimcoin \cite{misc9} is developed based on PoB.
	
	\textit{Proof of Inclusion (PoI):} The Merkle tree root \cite{merkle1987digital} in the block can be considered as a proof the inclusion of records which enables nodes to verify individual records without reviewing and comparing the entire chain. It can be said that the blocks of two nodes are verified and consistent if a copy of the blockchain has the same Merkel tree root for a block as another node’s copy of the blockchain. The Ethereum blockchain presented in \cite{wood2014ethereum} utilizes this PoI.
	
	\textit{Proof of Elapsed Time (PoET):} The PoET \cite{chen2017security} is designed by Intel for permissioned blockchain applications in order to address the challenges of expensive investment of energy in PoW. Based on the trusted enclave in Intel’s Software Guard Extension (SGX), the computational expensive works are replaced with the proof of elapsed time. PoET uses a trusted election model among the entire population of the validators, where it randomly chooses the next leader to publish the block. The validators in network request for a random wait time from their enclaves. The validator having the shortest waiting time for a particular block is elected as the leader, and it needs to wait until after the waiting time had expired to publish the new block. The trust is established in the hardware that produces the time. The Hyperledger Sawtooth \cite{misc10} is based on PoET.
	
	\textit{Proof of Authority (PoA):} The PoA \cite{de2018pbft} is designed particularly for permissioned blockchain. According to the mechanism, before becoming an authority to publish a block, the participant has to confirm its identity in the network. Unlike PoS, instead of having some coins/other assets, PoA considers a participant’s identity as a stake. Moreover, it is assumed that the authorities are pre-selected and trusted to publish a block. Also, it is convenient to detect the malicious authorities and inform about the malicious activities to other nodes. The Parity Ethereum \cite{misc11} is developed based on PoA.
	
	\textit{Practical Byzantine Fault Tolerance (PBFT):} The PBFT \cite{castro1999practical} provides a solution to the Byzantine Generals Problems \cite{lamport1982byzantine} for the asynchronous environment. PBFT works on the assumption that at least two-thirds of the total number of nodes are honest. It involves the following phases.
	
	\begin{enumerate}[label=\roman*.]
	\item A primary node is selected to become a leader in order to create and validate a block. The primary node can be changed by the rest of the nodes in the network, and the selection is also supported by more than two-thirds of all nodes.
	\item After receiving a request from the user, the leader generates a new block which is considered as a candidate block.
	\item The leader broadcasts the block to other nodes who are able to participate in consensus for verification as well as auditing.
	\item After receiving, each node audits the block data and broadcasts the results with a hash to other nodes. The audit results are compared by the nodes with others.
	\item The nodes reach a consensus on the candidate block and send a replay back to the leader which consists of audit and comparison results.
	\item Once the leader receives the results from at least two-thirds of the nodes agreed on that candidate block, the leader can finalize the block to include in the chain.
	\end{enumerate}
		
	Besides these aforementioned popular consensus mechanisms, some others include Algorand \cite{gilad2017algorand}, RAFT \cite{ongaro2014search}, Seive \cite{cachin2016non}, Tendermint \cite{kwon2014tendermint, buchman2016tendermint}, Ripple \cite{schwartz2014ripple}, Stellar \cite{mazieres2015stellar}, Proof of Space \cite{dziembowski2015proofs, park2018spacemint}, Proof of Importance \cite{misc12}, Proof of Exercise \cite{shoker2017sustainable}, and so on. However, the components of a block and consensus algorithms may vary and solely depend on the specifications of the use-cases of blockchain.

\section{Blockchain in Smart Grid}
In this section, we mainly present the potential opportunities of applying blockchain in smart grid. As such, we first discuss in detail what the future smart grid system will be. We next describe the blockchain features as well as the security, privacy \& trust objectives that can be addressed by blockchain in order to show how these will ultimately motivate to apply blockchain in smart grid.

	\subsection{Moving towards Decentralized Smart Grid System}
	As discussed in the introduction section that the smart grid concept has been represented a new grid infrastructure that uses digital computation and communication technologies to transform and modernize the conventional legacy grid into more accurate, efficient, and intelligent energy access and delivery network. These transformation and modernization have occurred due to the aggressive climate change and the necessity of sustainable energy sources. The ultimate aim of these transformation and modernization is to reform the energy landscape by integrating and utilizing more renewable \& distributed energy resources and lowering down the dependencies of fossil fuel-based generations. While the conventional legacy grid serves consumers through long-distance transmission lines, the smart grid paradigm brings the producers and consumers closer to each other by deploying independent distributed producers of renewable energy.
	
	Most recently, the energy Internet (EI) concept \cite{al2019iot, saleem2019internet, kabalci2019internet, wang2017survey, yijia2018comprehensive, dong2014smart, chen2019internet, hussain2019emerging} is introduced which is defined as the upgraded version of smart grid system. The EI is characterized by Internet technologies to develop the next-generation of smart grid by integrating information, energy, and economics. The EI is aimed to provide a great opportunity to facilitate the seamless integration of diverse clean and renewable energies with the grid and also, provide more interactions among various elements of the power grid to develop a fully autonomous and intelligent energy network. The key idea behind the EI is to be capably shared both energy and information similar to data sharing on the Internet. Here, the usual elements include traditional generation units, micro-grids, distributed energy resources (DERs), community-generated energy networks, energy storage units, electric vehicles (EV), vehicle to grid (V2G), cyber-physical systems, prosumers, service providers, and energy markets.
	
	Though the smart grid and energy Internet are aimed to make able to adapt both distributed generations and centralized energy generations, one key drawback of this current design is possessing centralized topology where energy generations, transportations \& delivery network, and markets are somehow dependent on centralized or intermediary entities. In this centralized system, the elements of the smart grid interact and communicate with centralized entities that can monitor, collect, and process data and support all elements with appropriate control signals. Moreover, the energy transmission is done usually over a long-distance network to deliver energies to the end-users through the distribution network. Unfortunately, due to the penetration of renewable energies as well as the continuously booming number of elements, the current design of smart grid system raises some concerns. Such concerns include scalability, expandability, heavy computational and communicational burdens, availability attacks, and incapable of controlling future power systems which will consist of a large number of components.
	
	As such, transforming into the decentralized system is a trend in smart grid to bring more dynamic, intelligent and proactive features. The grid infrastructure itself is also undergoing an adaptation and moving toward a fully automated network having decentralized topologies in order to increase the interactions among all components of smart grid systems in a dynamic way. The connectivity and accessibility that EI offers additionally reach a higher level of economical, efficient, and reliable operation of the smart grid system. Table III represents a brief comparison between the smart grid and the EI-enabled future decentralized smart grid.

\begin{table}[ht]
\centering
\caption{A Brief Comparison between the Smart and Future Decentralized Smart Grid}
\begin{tabular}{m{4cm}|m{4cm}}

\hline \hline
Smart Grid			& Future Decentralized Smart Grid \\
\hline
Transformed into utilizing more renewable energies and integrating with centralized grids		& Moving towards building a decentralized system by integrating various distributed energy resources \\
\hline
Generally, focus on the integration of advanced sensing and control technologies into traditional grid			& Basically, refer to real-time monitoring, auto-adjust controlling and optimization \\
\hline
Relied on intermediaries and centralized markets			& Support a number of users to generate their own energy and share surpluses through peer-to-peer \\
\hline
Utilized advanced communication technologies			& Dominated by energy Internet to realize Internet-like seamless energy and information sharing \\
\hline
Come up with bi-directional communications			& Support advanced plug-and-play functionalities \\
\hline
Heavy computational and communicational costs			& The costs are distributed among the entities over the network \\
\hline
Less option to expand the network			& Have the option to expand fast and large number of connectivity \\
\hline
Would be affected by a single point of failure			& Have resiliency against single-point of failure \\
\hline
Integrated with only electric energy networks			& Integrated with other energy networks as well \\
\hline
Dependent always on the regional system control			& Allows the smooth access of massive distributed energy resources \\
\hline

\end{tabular}
\end{table}	
	
	\subsection{Motivations of Applying Blockchain in SG Paradigm}
	The security, privacy, and trust are the key concerns to every system. In same connection, the future smart grid system should also have some level of security \cite{el2018cyber, weerakkody2019challenges, hossain2019application} such as (i) ensuring that any unauthorized entity cannot obtain any information, (ii) ensuring proper cryptographic mechanisms, (iii) preventing unauthorized entities from modifying the information, (iv) refraining access by any entity without permission, (v) ensuring access by those with the rights and privileges, (vi) providing evidence that an entity performed a specific action so that the entity cannot deny what it has done, (vii) developing a fault-tolerant network having resistance against availability attacks, (viii) making more efficient monitoring, (ix) leveraging advanced privacy-preserving techniques to protect information disclosure, and (x) increasing the trust, transparency and democracy among all the entities.
	
	In \cite{nakamoto2008bitcoin}, Nakamato solves the problem of establishing trust in a distributed system by introducing a novel consensus mechanism that makes Bitcoin the most successful blockchain application so far. This is due to not only using this consensus mechanism but also utilizing other techniques such as cryptographically protected data structure, digital signature technique, time-stamp, and rewarding scheme. Specifically, in blockchain applications, the consensus mechanisms are usually utilized to establish trust only. On the other hand, the different kinds of cryptographic techniques are also utilized to solve mainly the basic security requirements including confidentiality, integrity, authentication, authorization, non-repudiation as well as privacy. Furthermore, other techniques are utilized to support blockchain technology which we have already mentioned in Section II. Though the concept of consensus mechanism and blockchain is initially introduced in cryptocurrency applications, it is not required to build a cryptocurrency to develop a blockchain-based decentralized system.
	
	Currently, most of the solutions are built on centralized models where smart grid components are dependent on either centralized platforms or intermediaries to get services like billing, monitoring, bidding, and energy trading, etc. Though these solutions are matured and working rightly, several challenging issues are associated with the current smart grid system. Moreover, we already mentioned previously that the smart grid is facilitating the integration of a large number of EVs, DERs, prosumers and cyber-physical systems. Thus, the grid topology itself is adapting and shifting from centralized topology to decentralized and fully automated network to allow greater interaction among the components. Also, the smart grid market is transforming to decentralized prosumers interactive network from centralized producer governing network with the help of EI concept.
	
	In this moving towards decentralized systems, applying blockchain presents an opportunity to facilitate this transformation due to its following features which makes it suitable to apply.
	
	\textit{Decentralization:} The blockchain network is usually maintained by different decentralized nodes through consensus protocols. This network can normally run in peer-to-peer manner without trusting a centralized trusted authority for authorization and maintenance.
	
	\textit{Scalability:} The nodes in the blockchain network are capable of scaling up the network as more and more nodes can join the network. This is mainly due to the decentralized nature of blockchain network which is maintained by a network of peers.

	\textit{Trustless but Secure:} Blockchain network is trustless but secure, as the nodes are not dependent on any trusted intermediary to communicate with each other and also, all records/transactions are secured by asymmetric cryptography. Unlike other systems, blockchain does not require trusting blindly certain entities.
	
	\textit{Immutability:} Since the blockchain technology utilizes cryptographic techniques and maintain a global ledger which is synchronized among the nodes, the contents inside the blocks cannot be altered unless the majorities become malicious.
	
	\textit{Transparency and Auditability:} Blockchain network is highly transparent by its structure as the nodes of the network are able to verify the authenticity of the records and have an assurance that the blocks are not altered. Moreover, this transparency makes the blocks auditable to any node on the network by opening all the records to everyone.
	
	\textit{Resiliency:} Blockchain technology enables a resilient and fault-tolerant network in which any fault or malicious activities can be identified and recovered easily. This resiliency comes with the decentralization of the architecture with no single point of failure and also, storing the entire chain by all nodes in their premises.
	
	\textit{Secure Script Deployment:} One more important benefit comes with the immutability and decentralization is secure script deployment inside the blockchain. In blockchain context, it is also called smart contract. Usually, for smart contract, the contracts are stored on the blockchain. These contracts can execute independently and automatically based on some predefined criteria without human intervention, broker, and any central authorization.

	Thus, with these features as mentioned above along with the cutting-edge cryptographic security benefits, blockchain can be a promising alternative to the conventional centralized systems to improve security, privacy, and trust while assisting in removing the barriers to become a more decentralized and resilient system. Before discussing the blockchain contributions in the context of smart grid, we present in Table IV the common security, privacy, and trust objectives but they are necessary for smart grid as well, and also, how blockchain can achieve these objectives.

\begin{table}[ht]
\centering
\caption{Summary of the Common Security, Privacy and Trust Objectives and the Descriptions of How Blockchain can Address}
\begin{tabular}{m{1.7cm}|m{6.3cm}}

\hline \hline
Objective			& How Blockchain can Achieve \\
\hline
Confidentiality		& Usually, the records are not encrypted in public blockchain; Cryptographic techniques \\
\hline
Integrity			& Cryptographically protected data structure by hash function, Markel tree, nonce (numbers used once) and time-stamps; Manipulated records can be detected and prevented decentralized access \\
\hline
Authentication		& Signed records inside the blocks by users’ individual private keys so that it can be verified that only the valid user sent it \\
\hline
Auditability		& Publicly available records/transactions in public blockchain \\
\hline
Authorization and Access Control		& User-defined authorization and access control relied on smart contract; Attribute certificates \\
\hline
Privacy			& Pseudo-anonymization by using hash functions to keep secret identities, Zero-knowledge proof \\
\hline
Trust			& Consensus algorithms, Trust is not placed on central entities/intermediaries rather it is distributed among the entities in the network \\ 
\hline
Transparency		& Complete transparency by maintaining an immutable distributed ledger including all records, transactions, events, and logs \\
\hline
Availability		& Distributed architecture with allowing multiple entities to establish connections with others and to replicate full copy of the blockchain \\
\hline
Automaticity		& Blockchain and smart contract offers automaticity where the entities can communicate and exchange values in peer-to-peer way by blockchain and execute actions automatically by smart contract \\
\hline

\end{tabular}
\end{table}

\section{Blockchain Contributions in Smart Grid}
In this section, we concentrate on the blockchain contributions towards smart grid by presenting several recent works. In the following, we first explore the blockchain solutions on typical smart grid security, privacy, and trust of smart grid. We then focus on the blockchain contributions on smart grid areas such as advanced metering infrastructure, decentralized energy trading \& market, energy cyber-physical system, management of EVs \& its charging units, and finally, microgrid. Such blockchain applications are illustrated in Fig. 2. Moreover, in every section, before presenting the blockchain contributions, we discuss the challenges which can be possible to address with blockchain.

\begin{figure} [!t]
	\includegraphics[width=\linewidth]{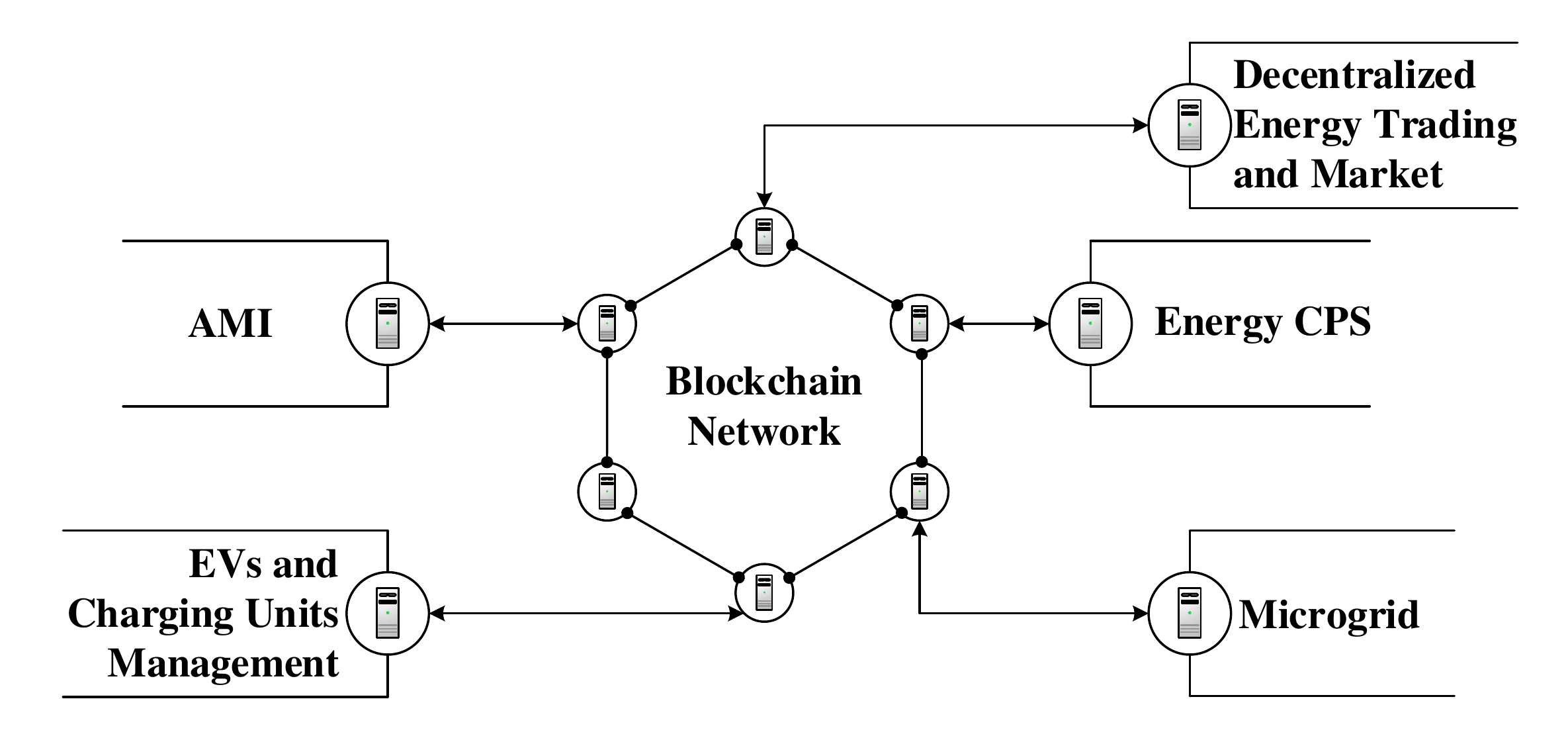}
	 \caption{Blockchain applications in smart grid domain that covered in this work.}
\end{figure}

	\subsection{Blockchain for Advanced Metering Infrastructure}
	With the introduction of advanced metering infrastructure (AMI), the utility companies, consumers, and producers in smart grid network can even more interact with each other through the automated and two-way communication supported smart meters. Compared with the traditional meters, smart meters are advanced meters which are able to collect the energy usage \& production, status, and diagnostic data in details. This data is often used for the purpose of billing, user appliance control, monitoring, and troubleshooting. However, these diverse data transfers are done through the wide area network and stored in traditional centralized storage system or cloud. The existence of the centralized system may involve its inherent issues like potential risks of modifications, privacy leakages, and single point of failure. Besides, more connections with a centralized system may also make scalability, availability, and delayed response problems. Apart from these, the smart meters and electric vehicles in smart grid system generate an enormous amount of payment records and energy usage data which are usually shared with other entities for monitoring, billing, and trading purposes. However, in such a complicated system, this widespread data sharing introduces serious privacy risks since the data may be revealed considerable and sensitive information about identities, locations, energy usage \& generation patterns, energy profiles, charging, or discharging amounts by the middleman, intermediaries, and trusted third parties. Moreover, the trust issue is present among the centralized parties, producers, and consumers. Thus, producers and consumers may face some difficulties to accept fairness and transparency from centralized parties. How to develop a secure, privacy-preserving, and trusted decentralized AMI system is an important task. In this subsection, we summarize some related blockchain studies on AMI.
	
	The authors in \cite{mylrea2017blockchain} introduce a model where they explore blockchain along with smart contracts for smart grid resiliency and security. The contracts will act as an intermediary between energy consumers and producers to reduce cost and also, improve the transactions rate while improving the security of transactions. Once a transaction takes place, the smart meter connected with blockchain network will send the record to create a new block by including a timestamp for future verification purposes within the distributed ledger. The consumer can then be charged based upon the data which was recorded on the ledger. However, the lack of detailed discussion of technical aspects is one major concern of this work.
	
	In \cite{pop2018blockchain}, a model of demand-side management for smart energy grids is introduced to realize decentralization and autonomy. In this model, the blockchain is utilized to make decentralized, secure, and automated energy network so that all of its nodes will work independently without relying on centralized supervision and DSO control. In addition to this, it is utilized to store the energy consumption information in a tamper-proof manner in the blocks which can be collected from the smart meters. On the other hand, the smart contract as in Fig. 3 is presented to offer decentralized control, calculate the incentives or penalties, validate the demand response agreements, and apply the rules associated with making a balance between the energy demand and production in the power grid side. Finally, this model is validated and tested by building a prototype which is developed in the Ethereum blockchain platform with the help of energy consumption and production traces of UK buildings datasets. The results indicate that this model is enabled to adjust the demands timely in near real-time by performing the energy flexibility levels as well as able to validate all the demand response agreements. However, it is not clearly mentioned that how energy profile anonymity has been ensured in this public blockchain. By analyzing the publicly available transactions, there is a possibility to retrieve the user.

\begin{figure} [h]
	\includegraphics[width=\linewidth]{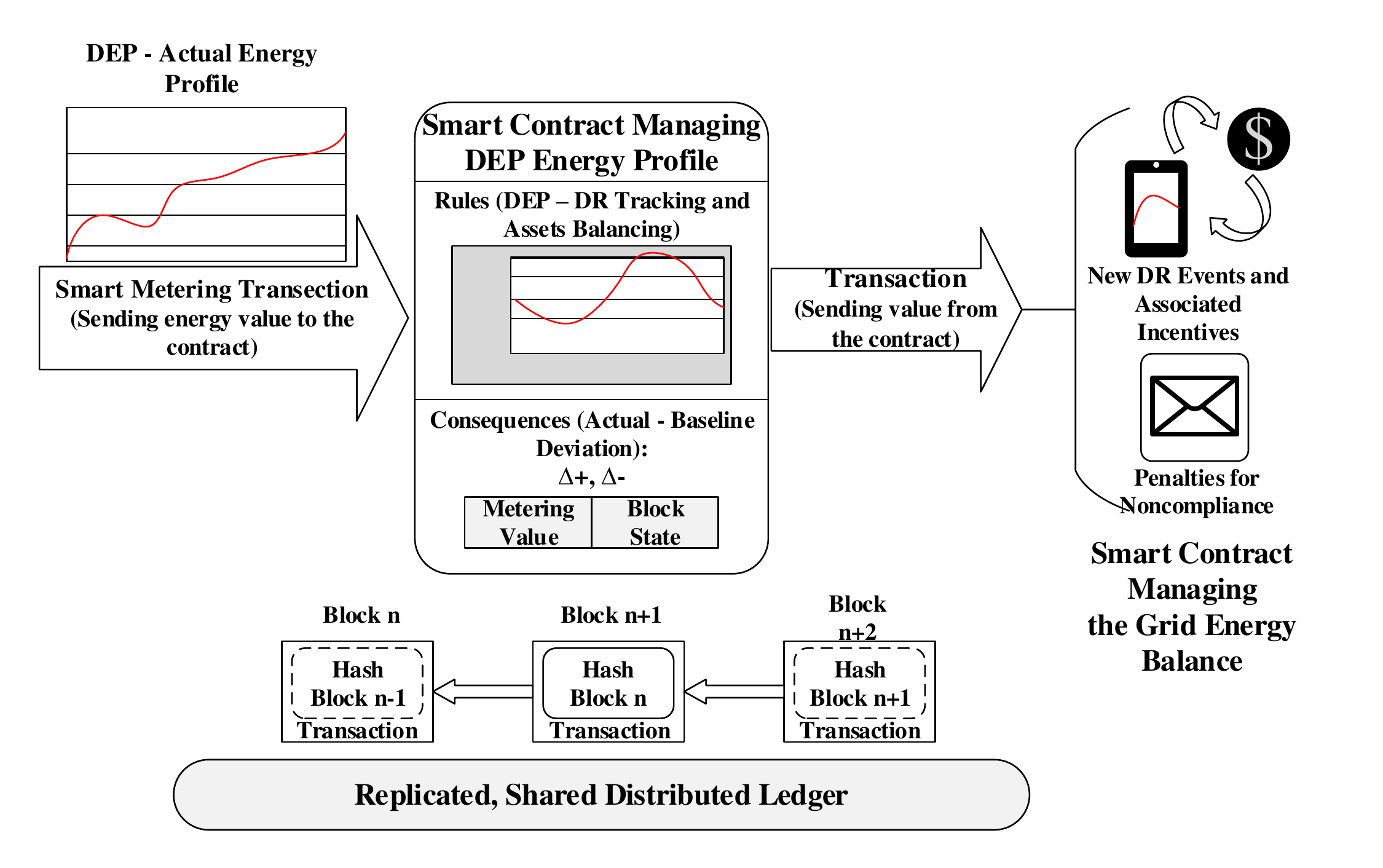}
	 \caption{The self-enforcing contract structure introduced in \cite{pop2018blockchain} for demand response tracking and asset balancing.}
\end{figure}

	The work in \cite{gai2019permissioned} introduces a blockchain and edge computing assisted approach to strengthen the functionality and energy security of smart grid network. The blockchain is utilized mainly to ensure the privacy of all participants and decentralized data storage in order to resist malicious activities within the communication channels and central data centers/clouds. This blockchain architecture is permissioned where three entities such as edge devices, super nodes, and smart contract servers as illustrated in Fig. 4 are introduced in order to ensure the correctness and trustworthiness within the blockchain network. Here, the edge devices are considered as typical nodes as like as contemporary blockchain system. On the other hand, the super nodes are a special type of nodes which are permissioned to select some devices from the edge devices to participate in consensus and voting process. Before asking to participate in the voting process, super nodes need to validate the identities of edge nodes through identity authorization and covert channel authorization techniques in order to make sure that the voting nodes are not malicious, and also, less likely to be compromised by the 51\% attack. However, the smart contract server nodes are responsible for implementing and attaching the contract script to the blocks. This smart contract contains the optimal strategy managed by edge devices for energy resource allocation to electricity users by considering energy consumption, latency and security. However, there will be a point of integrity concern, if the super node (SN) is compromised.
	
\begin{figure} [h]
	\includegraphics[width=\linewidth]{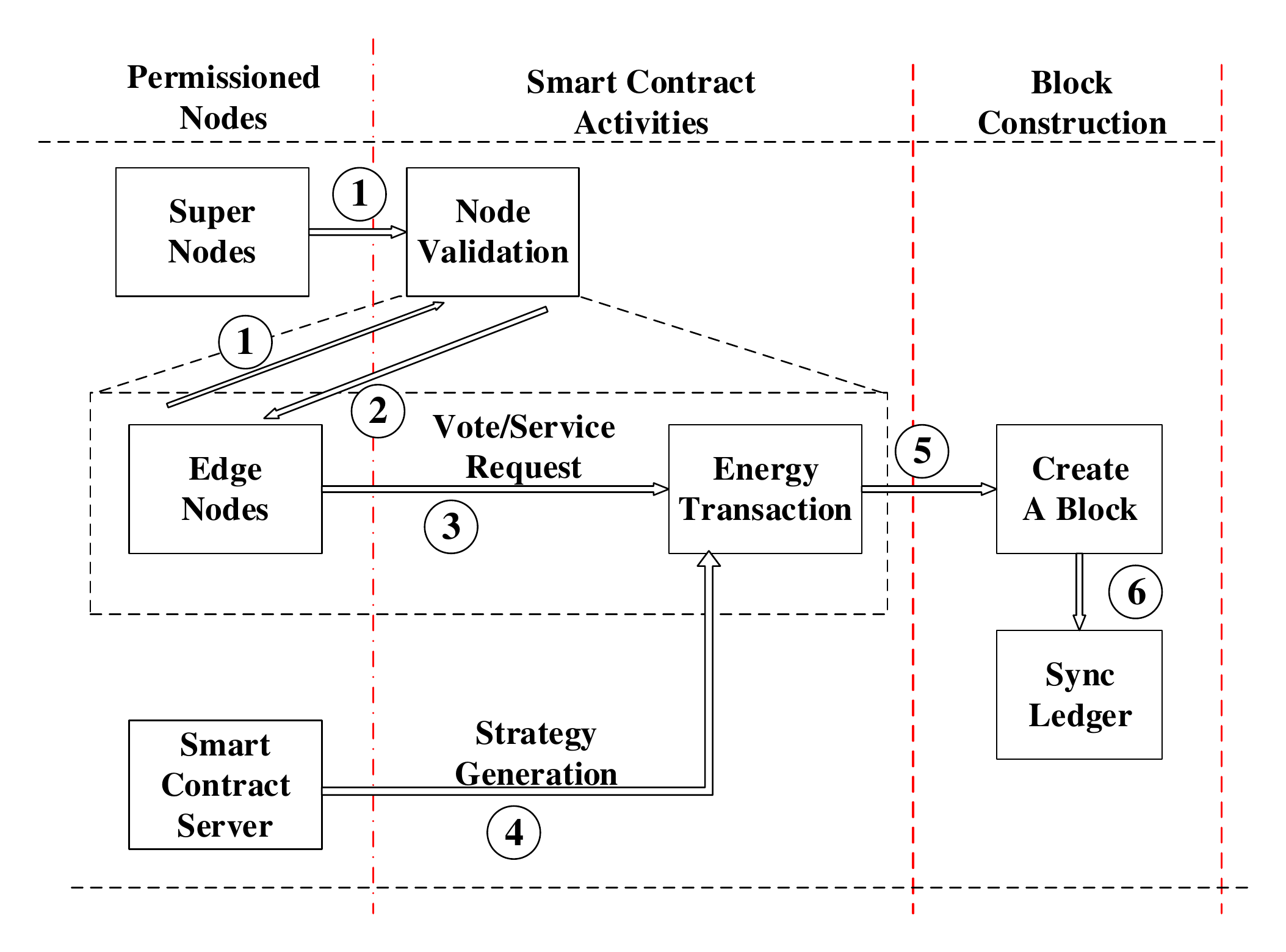}
	 \caption{The three main entities of proposed approach presented in \cite{gai2019permissioned} and their working relationships.}
\end{figure}
	
	The authors in \cite{tan2019privacy} introduce a secure and reliable energy scheduling model named PPES (Privacy-Preserving Energy Scheduling) for energy service companies (ESCO). Here, through the blockchain and smart contract, they address the growing privacy concerns of centralized ESCOs which may encompass financial and behavioral information which would cause privacy issues for the distributed energy market. Moreover, using a Lagrangian relaxation method, the proposed model is decomposed into some individual optimal scheduling problems. Afterward, the consensus and smart contract will solve the individual scheduling problems in the network in order to provide an overall reduction in energy costs while protecting privacy. Finally, the simulation and performance comparison show that the proposed model is much more efficient, reliable, and conducive than conventional models in achieving energy scheduling, information transparency, and energy trading. However, the scalability of this proposed model is not evaluated in this work.
	
	\subsection{Blockchain in Decentralized Energy Trading and Market}
	The bi-directional energy and information flow feature allow consumers to act as a producer and vice versa. Smart grid is expected to accommodate an increasing number of consumers, producers, and prosumers (producer + consumer) into distributed energy trading scenarios. As such, they should be able to trade their local generations or surplus energy from distributed sources such as microgrids, electric vehicles and energy storage units with each other in order to ensure some benefits like reducing load peaks, decreasing power loss in transmission, mitigating the burden of power grid in order to encourage green systems, and balancing energy supply and demand. Thus, it is necessary to integrate energy trading along with its necessary formalities which can offer bid handling, negotiation, and also, contract executions among the participants. Also, consumers and producers are allowed to trade energy with each other directly and seamlessly. This direct energy trading without involving any intermediaries can also enhance the benefits of all parties and is helpful for renewable energy deployments. However, in traditional methods, the consumers and producers can only participate in such trading formalities with each other indirectly through numerous intermediary third parties and retailers which will experience some potential issues and challenges. Consequently, it introduces high operational and regulatory costs which are ultimately transferred to consumers, producers, and prosumers. Furthermore, failure, compromised or malicious intermediaries lead to uncompetitive market, low transparency \& fairness and monopoly incentives, rewards as well as penalties. The salient properties of blockchain make an good tool to design a more decentralized and open energy market and trading.
	
	In \cite{li2017consortium}, Li et al. introduce an energy coin and peer-to-peer (P2P) energy trading system to ensure the security and decentralization of energy trading in different scenarios including energy harvesting, microgrids, and vehicle-to-grid networks using the concept of consortium blockchain technology, credit-based payment scheme, and Stackelberg game theory. Fig. 5 shows the proposed system with its four entities and their corresponding processing steps. The credit-based payment system is introduced in order to overcome the problem of transaction confirmation delay which is very likely in PoW based Bitcoin. Under this scheme, the peer nodes can apply for energy coin loans under their credit values from the credit banks so that they can make fast and efficient payments unlike Bitcoin. The Stackelberg game theory was utilized to propose an optimal loan pricing strategy for this scheme to maximize the economic benefits of credit banks. However, in this work, a formal proof on double-spending attack is not discussed, and the prototype of the proposed energy blockchain is not implemented.
	
	Inspired and built upon by Bitcoin, authors in \cite{aitzhan2016security} introduce a token-based decentralized system named PriWatt. The aim of this PriWatt is to address the problems of ensuring security of transactions as well as privacy of user identities in smart grid energy trading system. This system consists of blockchain-assisted smart contract, multi-signatures, and anonymous encrypted messaging streams. Through the agreements written within the smart contract, PriWatt allows the buyers and sellers to handle complex bidding and negotiation of energy prices while preventing malicious activities. In order to do bidding and negotiation anonymously, anonymous messaging stream technique is utilized. The multi-signature scheme is used to ensure protection against theft, whereas to validate a transaction, minimum multiple parties need to sign that transaction. Moreover, for consensus, they also use PoW like bitcoin to avoid Byzantine failures and double-spending attacks. However, there is no detailed discussion on which nodes are the PoW miners, how PoW will be performed by miners, and what will be the rewards after successfully mining.
	
	The authors in \cite{zheng2018smart} propose a novel power trading approach for smart grid to ensure the efficiency, flexibility and also, protection of users’ data privacy which is developed by utilizing consortium blockchain, smart contract, PoS, pseudonyms, and cryptographic mechanisms. This proposal puts forward the use of PoS instead of expensive PoW within their network while also encrypting the users’ data collected by the sensors using cryptographic techniques to upload to the authorized nodes. The smart contract implemented in Ethereum was developed to enforce data access rights restrictions and transaction transparency. Moreover, the pseudonyms were used to ensure the privacy of the users by hiding their real identities. However, this work is lack of practical implementation and evaluation of communications overheads and energy consumption.
	
\begin{figure} [h]
	\includegraphics[width=\linewidth]{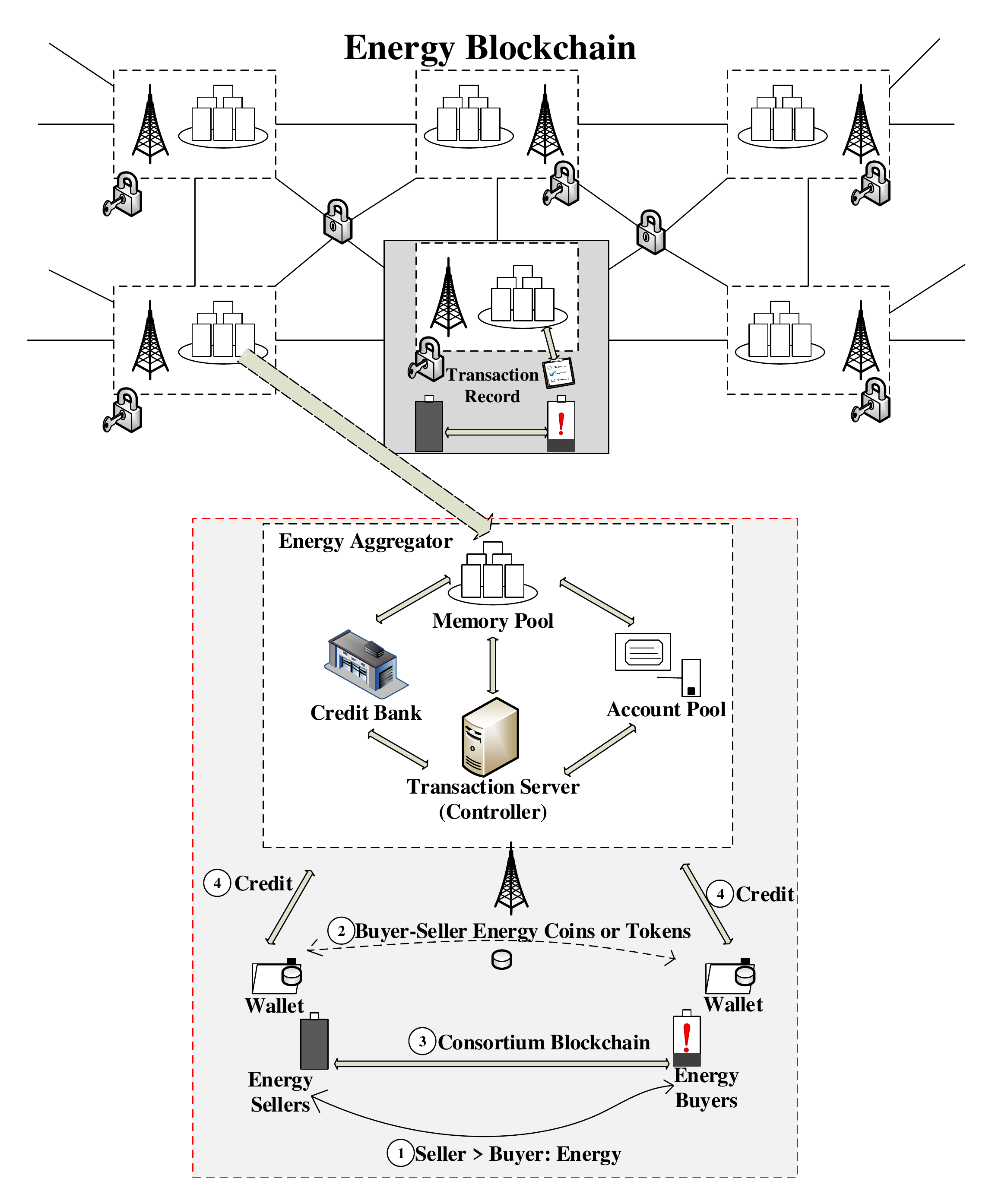}
	 \caption{Overview of consortium Blockchain-based secure energy trading system introduced in  \cite{li2017consortium}.}
\end{figure}
	
	Garg et al. \cite{garg2019efficient} introduce a blockchain and elliptic curve cryptography (ECC) assisted hierarchical authentication mechanism to ensure security \& privacy in energy trading for a distributed V2G environment. The systematic diagram of this proposed scheme is illustrated in Fig. 6 which includes its components of the considered ecosystem and working procedures. Within this mechanism, the blockchain is used to execute the transactions, whereas, the ECC is utilized for the purpose of hierarchical authentication. Here, the aim of this hierarchical authentication mechanism is to ensure the anonymity of Electric vehicles (EVs) and, also, offer mutual authentication in-between the communicating parties in V2G such as EVs, aggregators and charging stations. However, threat model is not discussed properly, and the proposed mechanism is not been practically evaluated.
	
	In \cite{wang2019energy1}, a blockchain-based crowdsourced energy system (CES) framework and operational algorithms are presented which facilitate P2P energy sharing at the distribution level. This framework involves various tasks such as charging/discharging EVs, deferring electric loads, and connecting with renewable energies. With the help of distributed blockchain implementations along with smart meters, these tasks are made automated. The ultimate aim of this CES framework is to address the real-time demand shortage as well as surplus issues in the network. On the other hand, the operational algorithm has two steps. The first step is developed to manage the bulk of grid-operation by focusing on the day-ahead scheduling of production and controllable DERs, whereas, the second step helps to balance hour-ahead energy deficit/surplus through incentives. These algorithms facilitate P2P energy sharing which results in a systematic way to regulate the distribution network. At the same time, it motivates and stimulates the crowdsources to contribute to the distributed network ecosystem such as microgrids. Finally, the CES framework and algorithms are prototyped within the Hyperledger fabric platform. However, it is not mentioned that how to address the attacks from malicious crowdsourcers, market stakeholders, and outsiders.
	
\begin{figure} [h]
	\includegraphics[width=\linewidth]{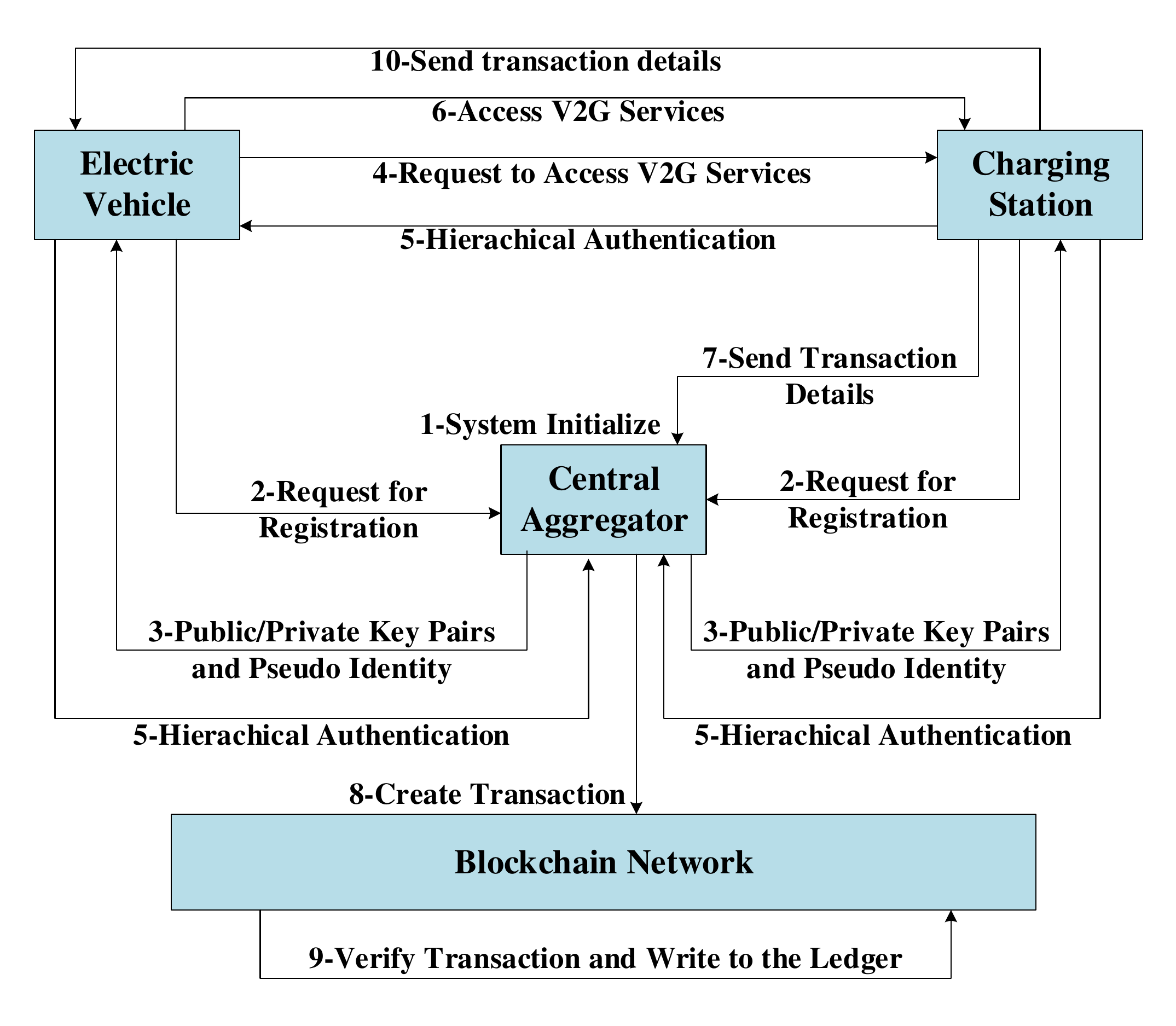}
	 \caption{The systematic diagram of the proposed scheme presented in \cite{garg2019efficient}.}
\end{figure}

	\subsection{Use of Blockchain to Monitor, Measure, and Control}
	The present smart grid cyber-physical system (CPS) is mainly built on centralized supervisory control and data acquisition (SCADA) system which is interconnected with various elements such as MTUs, RTUs, PMUs and a number of sensors on a hierarchical manner. The SCADA system is widely utilized to monitor and control the power grids. Integrated with the Internet, the SCADA system will enable large-scale distributed monitoring, measurement, and control in an improved way. The IoT smart devices, sensors, and PMUs usually collect status information of power devices and share with MTUs through RTUs, whereas the MTUs are considered as central repositories and control centers. The smart grid system utilizes these fine-grained measurements among CPS elements, different grid operators, suppliers, and consumers which enable intelligent control, wide area monitoring, and governance to better manage their grid’s safety, stability, reliability as well as monitoring power theft and loss. However, cyber-attacks can be launched in different ways by malicious attackers or insider such as altering data within central controllers, launch availability attack, and injecting bad data trough sensors and PMUs. As a result, the attacker is able to take over control channels and also, generate malicious commands. In the decentralized smart grid system, the blockchain offers new opportunities to monitor, measure and control.
	
	The authors of \cite{maw2019ics} focus on data security in the industrial control system (ICS). A blockchain-based architecture called ICS-BlockOpS is introduced to increase the security of plant operational data. This architecture is mainly developed to address two major issues in ICS such as immutability and redundancy by utilizing blockchain technology. The tamper-proof nature of blockchain is able to ensure data immutability. On the other hand, to provide data redundancy, a blockchain assisted efficient replication mechanism is also presented to ensure data redundancy which is inspired originally by the Hadoop Distributed File System (HDFS). However, there are no discussions on how to address false data injection by malicious or compromised nodes, and how resource-limited sensors and actuators will work in blockchain network.
	
	Ref. \cite{gao2018gridmonitoring} presents a blockchain and smart contract based monitoring system on smart grid to ensure energy consumption transparency and security. This blockchain network consists of three different kinds of nodes according to their responsibilities such as smart meters, consensus nodes, and utility companies. Usually, the smart meters send digital measured energy data from consumers to the blockchain network. On the other hand, the consensus nodes are responsible for handling the energy consumption records, keeping individual subscription details provided by utility companies, validating to create new blocks, and broadcasting them to add to the main chain. However, before creating new blocks, these nodes make temporary forms for individual users which include meter IDs and other consumer information. Later, these forms will be converted into blocks once audited and accepted by consensus nodes. Here, the blockchain is adopted to create an immutable data record system for protecting smart meter data records against manipulation from both consumer and utility company sides. Moreover, the smart contract is utilized to increase transparency by setting some rules to identify malicious usage of electric power, detect malicious manipulation of usage data and enforce penalties. However, this work lacks of practical evaluation for security measures and performance efficiencies.
	
	In \cite{wan2019blockchain}, the authors introduce a blockchain-assisted partially decentralized cyber-physical system architecture in order to address the issues of traditional cloud-based architecture. This architecture is comprised of five layers as shown in Fig. 7. The first layer composes different types of sensors and computing devices. The responsibilities of computing devices are to collect and preprocess the data from sensors. The second layer does cryptographic operations on the data, generates blocks, and then, records them in the distributed ledger. The function of the third layer is to store the entire blockchain in a distributed and synchronized manner. To make all these aforesaid layers complement each other effectively, another layer as the fourth layer is introduced. It involves the underlying implementation technologies including distributed algorithms and data storage technology in order to connect each layer. The last layer provides real-time monitoring and failure prediction like services to the users. Lastly, the proposed architecture is implemented in an automatic production platform. Results from the experiments depict that in comparison to the traditional architecture, the proposed architecture provides better security and privacy. However, this paper lacks of discussion on which nodes are responsible for performing PoW and what is the responsibilities of lightweight nodes.
	
\begin{figure} [h]
	\includegraphics[width=\linewidth]{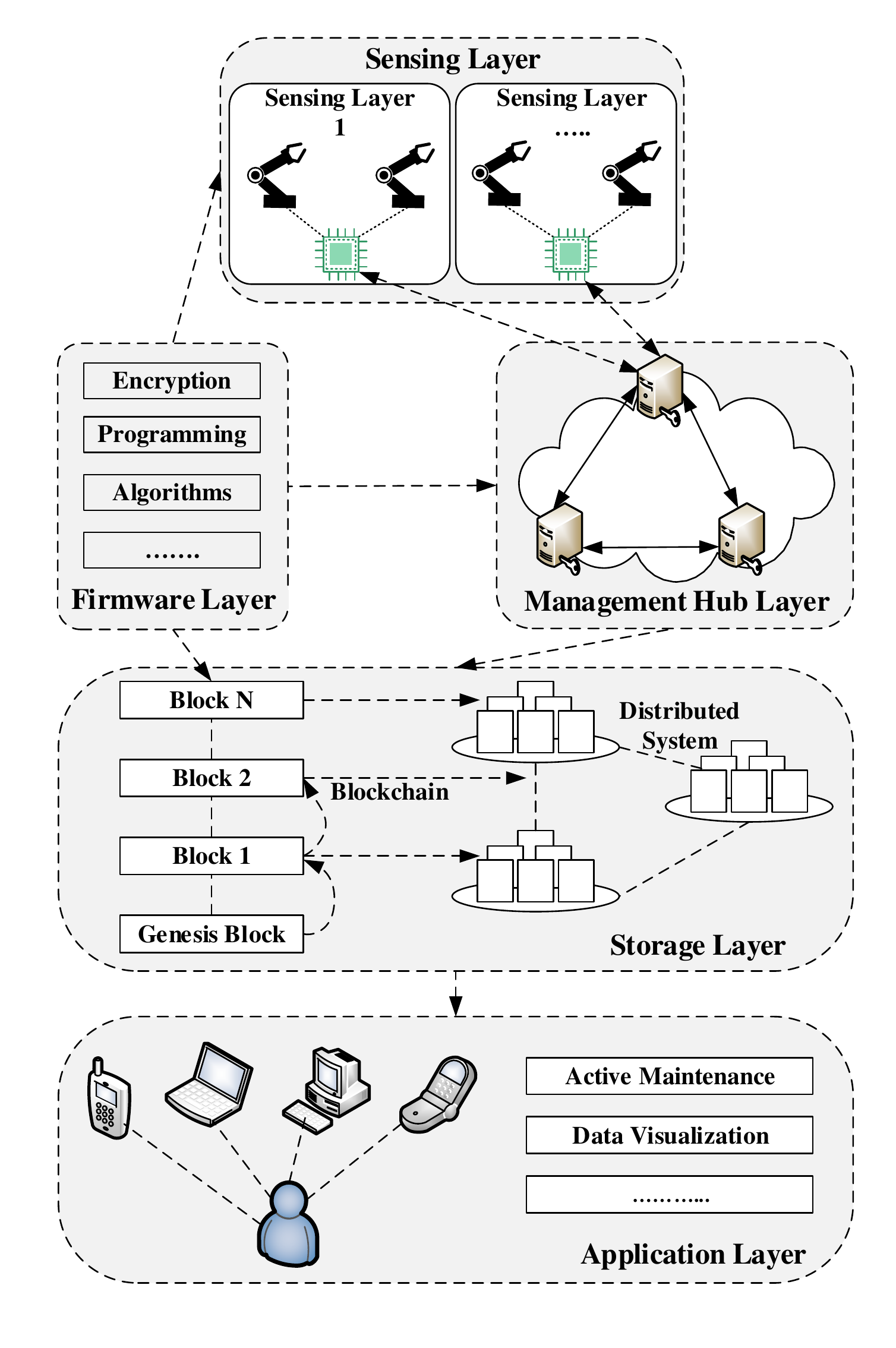}
	 \caption{The Blockchain-assisted architecture presented in \cite{wan2019blockchain}.}
\end{figure}
	
	\subsection{Blockchain on EVs and Charging Units Management}
	Electric vehicles are regarded as one of the cornerstones of the future smart grid which can act as energy storage devices and are able to exchange energy with power grid, charging stations and other neighboring EVs in a P2P manner. This leads to three scenarios, i.e., Vehicle-to-Vehicle (V2V), Vehicle-to-Grid (V2G), and Grid-to-Vehicle (G2V). In this way, EVs can play a key role in contributing effective demand response, enhance grid resilience and reduce the load peaks by energy charging and discharging operations. Different from the present situation, delivering and managing a vast number of EVs are expected from EI. However, frequent two-way power \& data communications within aforementioned three scenarios, short communications range, and EVs mobility can introduce new security and privacy issues. Moreover, high-level penetration of uncoordinated charging may lead to grid overloading. As a result, it is becoming open challenges to integrate a large number of distributed EVs, optimal charging schedule along with individual energy preferences, and develop a transparent charging coordination mechanism. Thus, it is crucial to develop decentralized and transparent EVs and charging management mechanisms. To develop such mechanisms, the blockchain is used in several literatures as discussed below where the researchers try to enhance the EVs and its charging management.
	
	The authors of \cite{su2018secure} focus on building a smart contract empowered permissioned blockchain system in order to implement a secure EV charging framework integrated with renewable energy sources (RES) as well as smart grid in smart community (SC). To implement smart contract, the contract theory is utilized to design the optimal contracts and a novel energy allocation algorithm. These optimal contracts are made in-between the energy aggregator and EVs under it which will allow EVs to choose the source for energy consumption according to their individual preferences while maximizing both the operator’s and EVs' utilities. On the other hand, the energy allocation method is also introduced in order to allocate the limited energies from RES into EVs. Moreover, to achieve an efficient and fast consensus, this permissioned chain utilizes delegated Byzantine fault tolerance (DBFT) in where only the preselected EVs are allowed to participate in auditing and creating a new block. However, there are no discussion on who is responsible for validating the transactions and incentives for validators.
	
	The work \cite{baza2018blockchain} proposes a blockchain assisted charging coordination mechanism so that the energy storage units (ESUs) such as home batteries, EVs, and other storage units can get their charging demands in a transparent, reliable, decentralized manners from utility companies. The ESUs and utility companies are connected with the blockchain network. The utility companies are responsible for providing the maximum load profile information which is assigned to ESUs in each community. And, each ESU is responsible for sending a charging request to the blockchain network. This request contains its demand, state of charge and time to complete the charge. However, the smart contract is introduced in this mechanism so that the scripts inside it can schedule the charging requests in a decentralized, corrected and transparent way. Based on the aforesaid information and a Knapsack algorithm, the scripts will make a priority index for each ESUs which determines the ESUs with highest priorities can only charge in the present time slots, whereas the ESUs with lower priorities will be delayed to next time slots. Moreover, this mechanism preserves ESU identity information secret and defends from external malicious charging requests by using a number of certified pseudonyms for each ESU to use during charging requests. However, there is no discussion about how consensus mechanism will work.
	
	In \cite{huang2018lnsc}, the authors outline a security model for EV charging management to improve the security of EVs and make the present system into a decentralized one. This model is composed of the lightning network and blockchain. In the beginning, a lightning network is built-up where the EVs, charging piles, and operators get them registered with this network. The aim of this lightning network is to support the main blockchain network by establishing trust among the parties and also, assuring the security of both funds and payments. Afterward, various scheduling strategies are introduced to schedule the charging piles according to the policies of the operators as well as the requests from EVs. Once the EVs receive scheduling recommendations from the operator, they will go through authentication and authorization phases. In this way, the EVs complete the charging and the records will be written in the blocks. Finally, the authors evaluate this security model through an experiment using real EV traffic. The experimental results show that this model can effectively increase the security performance of energy trading between the EVs and charging piles. Also, it can easily incorporate present scheduling mechanisms. However, the security goals discussed in this paper are not directly relevant to blockchain-based system.
	
	Knirschet al. \cite{knirsch2018privacy} introduce a transparent, autonomous and privacy-preserving method where the EVs are allowed to find the cheapest and viable charging stations for them based on energy prices and the distances to the EVs. This method employs blockchain in the sense that when a bid request is made for a specific energy level, it is sent to the blockchain. Then, the blockchain will preserve the EVs’ identity privacy, hide their geographical location, make verifiable, and increase the transparency of bidding requests. Also, the charging stations connect with blockchain to keep their bid records for traffics based on the energy request. However, there are no discussion about how to make scalable for a large number of EVs, and how to handle the payment part in blockchain.	
	
	\subsection{Use of Blockchain in Microgrid}
	The microgrid, a grid paradigm, is becoming an integral part of the smart grid which is basically promoted by distributed energy resources (DERs) such as solar, wind and fuel cells. The microgrid is defined as a localized group of DERs, battery storage units, EVs, smart appliances and loads, where the generation units are usually located closed to the loads. Compared with the traditional centralized fossil fuel-based generations, the microgrid produces low voltage electricity on a small scale which has an aim to provide reliable electricity supply, reduce transmission losses, and utilize renewable sources. A microgrid network may host a large number of DERs and are connected to the grid in a distributed manner to inject available power to the grid. The plug-and-play microgrid integration is expected with the introduction of the EI-enabled smart grid. However, the high level of DERs penetration from microgrid network to the grid may lead to a power surplus that causes power grid unstable. Furthermore, this integration may result in some open challenging issues in energy trading and management. The energy management issues include congestion pricing, control, and optimality of dispatch. On the other hand, energy trading issues comprise improper incentive mechanisms to promote their adoptions and centralized \& monopoly markets. The blockchain can be utilized to address these challenges by developing decentralized microgrids.
	
	The authors in \cite{munsing2017blockchains} present a blockchain and smart contract assisted architecture to facilitate decentralized optimization and control of DERs in microgrid networks. In this architecture, the blockchain is considered to make the system decentralized by distributing the microgrid operators’ role across all entities and also, ensure fair energy trading without relying on a utility company or a microgrid operator. On the other hand, to coordinate the scheduling of DERs in the network, the smart contract is utilized. Moreover, a decentralized optimal power flow (OPF) model is also presented by employing the Alternating Direction Method of Multipliers (ADMM) technique in order to schedule the battery, shapeable, and deferrable electric loads in the distribution network. The local optimization step will be carried out by DERs, whereas the smart contract will serve as an ADMM coordinator. Finally, the optimal schedule will be kept on the blocks, and the payment transactions can be made securely and automatically. However, the authors did not explain how to build trust among non-trusted entities and also, how to address uncertain data.
	
	The authors in \cite{danzi2017distributed} focus on the problem of voltage regulations in microgrid network which is enormously influenced by the high DERs power penetration to the grid. Voltage regulation is important since overvoltage and undervoltage situations lead to severe damage. Overvoltage results in overheat which may damage power system infrastructure. On the other hand, undervoltage may result in the collapse of the system. The authors introduce a novel proportional-fairness control scheme for DERs in microgrid which is relied on blockchain as well as smart contract since the strategic operation can be helpful to allay voltage violations. Here, a group of DERs will be considered as voltage regulators and subsequently, reduce the penetration and sacrifice their revenue over control periods to balance the voltage regulations. The subset members will be selected dynamically based previous records of participations. Moreover, to provide incentives fairly among the members of the participant subset, the authors propose a principle based on the exchange of credits. Particularly, the participant DERs ask for a credit before going to participate as voltage regulators from those who are not participating. Later, these non-participants will be forced to engage in voltage regulation due to low credit. However, the role of blockchain is to make the current centralized system into a distributed one. Also, it stores the contracts, credit statues of DERs and history of the participants. On the other hand, to enforce the proportional fairness while receiving and paying the credits the smart contract is utilized. It also works as a distributed control authority trusted and operated by all DERs. Particularly, the DERs installed on the same distributed feeder make a contract among each other so that they can determine independently that which DERs will be able to participate as voltage regulators based on their stored credits. Afterward, it will confirm a fair shifting among the participant DERs to contribute to microgrid regulation. One limitation of this work is that any punishment mechanism for fraudulent transactions is not included.
	
	The work \cite{saxena2019blockchain} also addresses the voltage regulation problem where they introduce a blockchain-based transactive energy system (TES). This work is aimed to explore voltage regulation in almost the same context as the previous work \cite{danzi2017distributed}. But the study \cite{danzi2017distributed} does not pay any attention to provide punishments for fraudulent activities or incentives for voltage regulation services. With this proposed system, the distributed peers that possess renewable distributed generators (DGs) are allowed to ensure voltage regulation services to the power grid in exchange of economic payoff. Here, the concept of TES basically seeks to allow decentralized energy producers to carry out energy transactions and help in improving the performance of power system operation by providing computing services. Moreover, in order to improve the reliability as well as the efficiency of the power grid, TES attempts to integrate economic objectives and distributed control techniques. However, within this system, in the event of voltage violation, an initiating peer pleads for bidding from its neighboring peers, and a service contract is awarded to the most fitting peer. Additionally, a reputation rating is also observed for every peer as each successful contributor of voltage violation receives a positive reputation and vice versa. However, the peer negotiation process is enforced as smart contract. This process also utilizes a modified contract net protocol (CNP). After awarding a service contract to a peer, an enforcement stage is added in order to enhance the CNP. Here, the smart contract will substantiate the service contract by analyzing the most recent values of power measurement of peer DGs on the blockchain. This is done to confirm the successful contribution of voltage regulation. The implementation of this proposed system is demonstrated within a microgrid network which is separated into various zones. Every zone depicts a prototypical microgrid, and each peer is supposed to manage the voltage regulations in its own zone. Two sets of implementation results are presented in order to substantiate this proposed system as a proof of concept. The first result adopts a simulation model of a power distribution network. This enacts the proposed TES to make sure that the voltage violations can be allayed efficiently. Afterwards, the another result comprises an actual deployment the TES in a smaller microgrid infrastructure which is based at Vaughan in Canada. However, consensus mechanism is not discussed in this work.
	
	In \cite{wang2017novel}, Wang et al. address the traditional microgrid transaction management problems due to centralized trading system such as (i) trust issues among transaction center, buyers and sellers, (ii) fairness and transparency of transaction centers, and (iii) transactions security risk. Rather they introduce a decentralized energy transaction approach based on continuous double auction (CDA) mechanism and blockchain technology in order to support independent and direct P2P transactions between distributed generations and consumers in the microgrid energy market. In this approach, in the beginning, two trading parties present quotes to the market by following an adaptive aggressiveness (AA) technique to perform the transaction matching in the market. The AA technique is applied here so that the buyers and sellers can adjust the quotations dynamically according to the market information. On the other hand, the CDA allows multiple traders and buyers to bid in the market. Afterward, the CDA mechanism allows the parties to reach the market equilibrium quickly. Finally, the two parties, i.e., buyer and seller will complete the digital proof of energy trading by using multi-signature and blockchain. Here, the multi-signature technique helps to protect from any manipulation of a contract between buyer and seller, whereas, blockchain ensures the security of the transactions. However, it is not mentioned that how rich-rule problem of PoS will be addressed. The overall proposed structure of electricity transactions in microgrid is illustrated in Fig. 8.
	
	In \cite{sabounchi2017towards}, an electricity trading mechanism for microgrid prosumers (producer/consumer) is introduced to address the enegy market issues such as resilient management of real-time deregulation, non-optimality of dispatch, and congestion pricing. By leveraging blockchain, the prosumers are enabled to trade their locally generated energy with others directly in P2P manner. Apart from blockchain, the smart contract is utilized to incorporate auction models for energy trading into it. Afterward, to design the smart contract for the energy market, this mechanism adopts contract theory which will ensure real-time energy trading in deregulated and decentralized energy markets. In order to build a smart contract, they utilized the Ethereum platform. However, in this work, the incentive and punishment mechanisms are not discussed. The algorithm for the smart contract developed in this work is shown in Fig. 9.
	
\begin{figure} [h]
	\includegraphics[width=\linewidth]{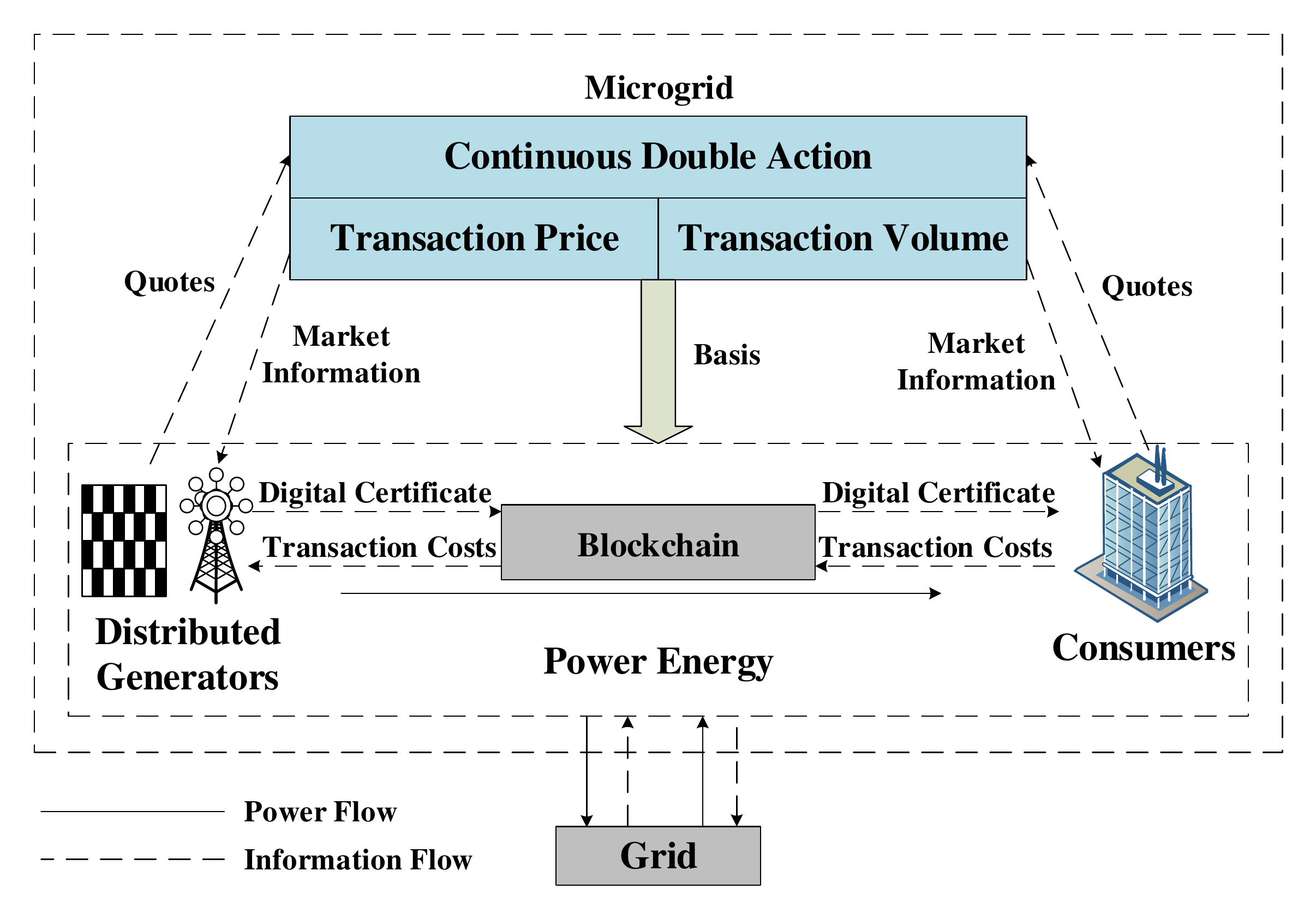}
	 \caption{The overall structure of the proposed electricity transactions in microgrid presented in \cite{wang2017novel}.}
\end{figure}

\begin{figure} [h]
	\includegraphics[width=\linewidth]{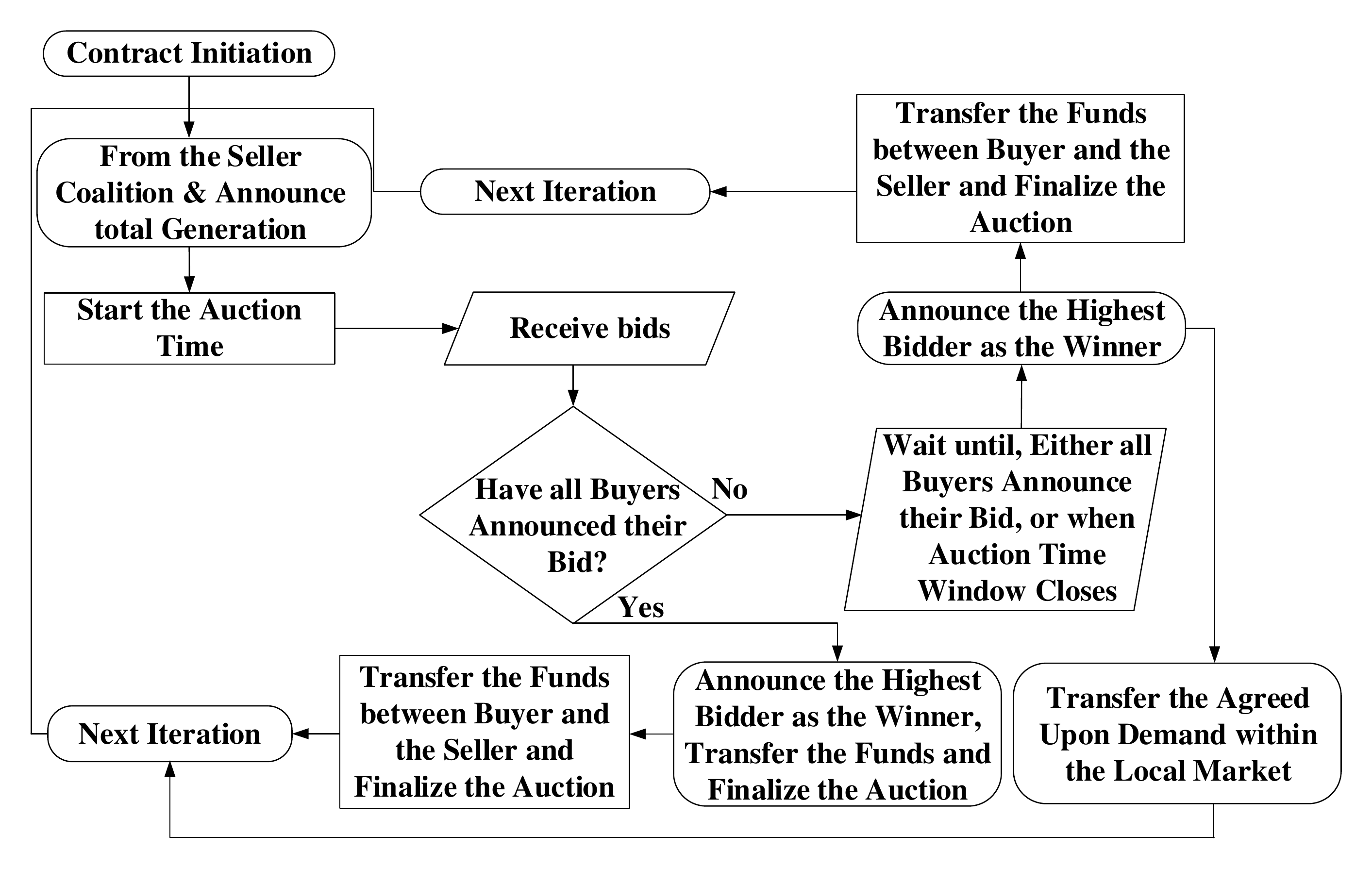}
	 \caption{The proposed algorithm for the smart contract of work \cite{sabounchi2017towards}.}
\end{figure}
	
	\textit{Summary:} In this section, we have introduced a number of blockchain contributions in the domain of smart grid where blockchain is utilized to record data, and smart contract is utilized to make automation. We have divided the contributions into five areas such as AMI, energy trading \& market, cyber-physical system, EVs \& their charging units management, and microgrid.  We also include individual limitations of these works. We present an overview of blockchain-enabled smart grid in Fig. 10. Then, we outline a summary of all blockchain-based solutions that we have discussed in this section in Table V. However, many of new solutions are still under development to come. From these limitations, we can say when designing a blockchain based solution for smart grid applications, the following issues should be considered such as (i) energy efficiency, (ii) balance among decentralization, security, privacy, efficiency, and scalability, (iii) security requirements and trust levels, (iv) specific targets/needs, (v) application scenarios, and (vi) practical validation.
	
\begin{figure*} [h]
    \centering
	\includegraphics[width=\linewidth]{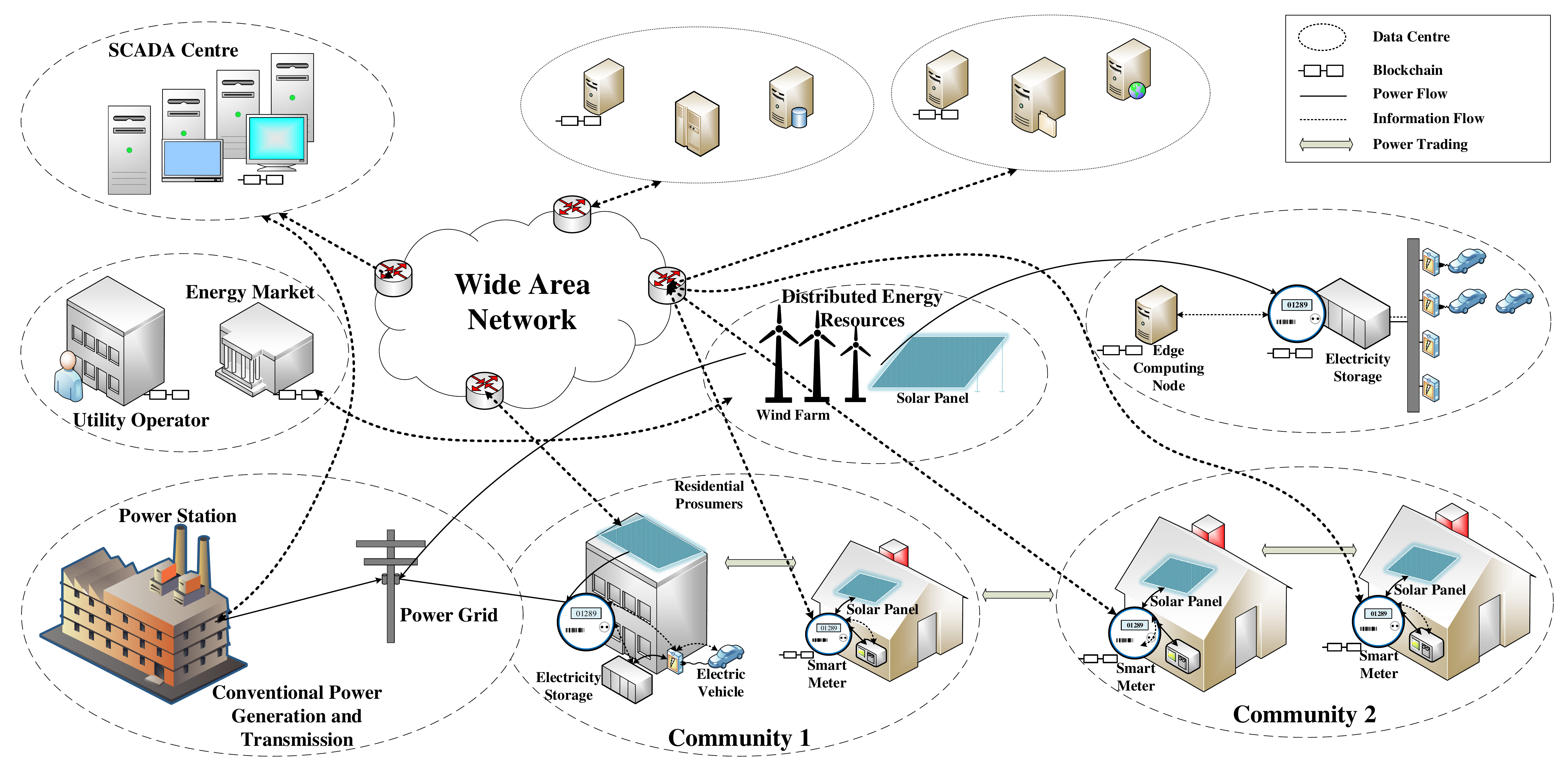}
	 \caption{An illustration of blockchain-enabled future smart grid.}
\end{figure*}

\begin{table*}[h!]
\centering
\caption{Summary of Blockchain-based Solutions in Smart Grid}
\begin{tabular}{m{.8cm}|m{6.5cm}|m{5.5cm}|m{3.5cm}}

\hline \hline
Work & Major Contribution & Technical Approach & Focused Application\\
\hline
\cite{mylrea2017blockchain}	&	Facilitating secure and fast energy exchange of DERs by means of blockchain-based AMI	&	Public blockchain and smart contract	&	Transactive energy applications\\
\hline
\cite{pop2018blockchain}	&	A blockchain-enabled distributed ledger for storing smart meter data (considered as energy transactions) to utilize in making a balance between energy demand and production	&	Public blockchain, Smart contract, Ethereum platform, and Proof of Stake (PoS)	&	Decentralized demand response programs management\\
\hline
\cite{gai2019permissioned}	&	A permissioned blockchain to ensure privacy and energy security (traceable and transparent energy usage) in smart grid		&	Group signature, covert channel authorization technique, smart contract, Permissioned blockchain, edge computing, pseudo names, and voting-based consensus		&	Traceable and transparent energy usage\\
\hline
\cite{tan2019privacy}	&	A blockchain-based privacy-preserving energy scheduling model for energy service companies	&	Lagrange relaxation algorithm, smart contract, and PoS consensus	&	Energy demand and supply information\\
\hline
\cite{li2017consortium}	&	A consortium-based energy blockchain	&	Credit-based payment system, consortium blockchain, and Stackelberg game theory	&	Peer-to-peer energy trading\\
\hline
\cite{aitzhan2016security}	&	A proof-of-concept deployment of blockchain-enabled secure energy transactions and privacy-preserving techniques to negotiate energy prices	&	Multi-signature, Anonymous Messaging Streams, PoW, Elliptic Curve Digital Signature Algorithm (ECDSA)	&	Decentralized energy trading and pricing\\
\hline
\cite{zheng2018smart}	&	A consortium blockchain-assisted efficient, flexible, and secure energy trading	&	PoS, consortium blockchain, smart contract	&	Smart grid power trading\\
\hline
\cite{garg2019efficient}	&	A blockchain-enabled hierarchical authentication mechanism for privacy-preserving energy transactions in V2G networks and rewarding to EVs	&	Elliptic curve cryptography (ECC), PBFT consensus mechanism	&	Energy trading in Vehicle to Grid (V2G) setup\\
\hline
\cite{wang2019energy1}	&	A blockchain-assisted operational model of crowdsourced energy system and energy trading	&	Smart contract, Redundant Byzantine Fault Tolerance (RBFT), permissioned blockchain	&	Crowdsourced energy system, P2P energy trading, and energy market\\
\hline
\cite{maw2019ics}	&	A blockchain-based architecture for industrial control system named ICS-BlockOpS to ensure operational data immutability, integrity, and redundancy	&	Smart contract and voting-based consensus	&	Interconnected cyber–physical systems (CPS)\\
\hline
\cite{gao2018gridmonitoring}	&	Applying blockchain to smart grid monitoring between electricity companies and consumers for data transparency	&	Smart contract and side-chain	&	Monitoring on smart grid\\
\hline
\cite{wan2019blockchain}	&	A blockchain-oriented partially decentralized architecture for more secure and reliable industrial CPS system and solving current limitations of cloud based system	&	Private blockchain, access control lists (ACL), PoW	&	Industrial CPS\\
\hline
\cite{su2018secure}	&	A blockchain-based solution coupled with contract theory to develop a secure electric vehicle charging framework including optimal scheduling algorithm and novel energy allocation in IoE	&	Contract theory, permissioned blockchain, reputation based DBFT consensus, and smart contract	&	Electric vehicles (EVs) charging services in smart community\\
\hline
\cite{baza2018blockchain}	&	A blockchain-based decentralized, transparent, and privacy-preserving charging coordination mechanism for ESUs such as batteries and EVs	&	Smart contract, Knapsack algorithm, partially blinded signatures	&	ESUs charging coordination in smart grid\\
\hline
\cite{huang2018lnsc}	&	A decentralized security model named LNSC based on blockchain to enhance the security of transactions between electric vehicles and charging stations	&	Lightning network, smart contract, elliptic curve cryptography	&	EVs and their charging pile management in IoE\\
\hline
\cite{knirsch2018privacy}	&	A blockchain-assisted automated and privacy-preserving protocol to search an optimum charging station relied on energy pricing as well as distance to the EVs	&	Smart contract 	&	EVs charging management\\
\hline
\cite{munsing2017blockchains}	&	A decentralized microgrid operational architecture builds on blockchain and alternating direction method of multipliers (ADMM) to address the monopoly price manipulation and privacy leakage problems by microgrid aggregators or operators	&	Smart contract and ADMM	&	Microgrid optimization and control\\
\hline
\cite{danzi2017distributed}	&	A blockchain-based proportional-fairness control framework to provide incentives to the distributed energy resources (DERs) for their contributions in voltage regulation in microgrid	&	Smart contract, PoW	&	Voltage control in microgrid\\
\hline
\cite{saxena2019blockchain}	&	A blockchain-based distributed voltage regulation algorithm for transactive energy system (TES)	&	Smart contract	&	Grid operation services for TES\\
\hline
\cite{wang2017novel}	&	A blockchain and continuous double auction (CDA) based decentralized microgrid electricity transactions mode to offer independent transactions between distributed generations (DG) and consumers	&	CDA, multi-signature, PoS	&	Electricity transactions in microgrid\\
\hline
\cite{sabounchi2017towards}	&	A decentralized transactive microgrid model	&	Smart contract and contract theory	&	Resilient networked microgrids\\
\hline
\end{tabular}
\end{table*}

\section{Practical Projects and Trials}
In order to promote the progression of the integration of blockchain and smart grid, several practical initiatives have been emerged most recently as trials, projects, and products. In this section, we present the key blockchain projects, industrial trials, and products focusing on different smart grid scenarios being deployed and published.

	\subsection{Cryptocurrency Initiatives}
		\subsubsection{SolarCoin}
		SolarCoin \cite{misc13} is an initiative to create and offer rewards for solar energy producers, who aims to provide incentives for a solar-powered planet. SolarCoin can be defined as digital tokens which are relied on blockchain. These digital tokens are maintaining at the rate of 1 SolarCoin per 1 MWh of produced solar energy. The purpose behind this is to enhance the encouragement of the development of the solar energy across the globe. This will result in the transition to a solar-based economy from a fossil fuel-based economy.
	
		SolarCoin compensates for the cost of electricity which enables solar installations to be paid off expeditiously. The solar energy producers are granted this SolarCoin freely. SolarCoin can be received by anyone who produces solar power. In order to register their solar installations, the solar energy producers of any size can freely submit a claim to any of SolarCoin affiliates. The claimants then download a free SolarCoin Wallet in order to create a receiving address. This address serves like a bank account that is shared with the affiliate along with some solar facility data and documentation. The SolarCoins are then sent to the claimant’s wallet by the SolarCoin Foundation at a rate of 1 Solar Coin per 1 MWh of validated electricity production. It is totally up to the claimants to save, exchange or spend the SolarCoins wherever they want. Moreover, SolarCoin can be commutated for government currencies on cryptocurrency exchanges as well as can be spent on businesses that accept them. It can also be spent or traded for goods or services and Bitcoin or other major cryptocurrencies. Simultaneously, SolarCoin can also be used to pay for other digital and foreign currencies.
	
		The PoST (Proof of Stake Time) consensus mechanism originally evolved in the VeriCoin approach \cite{misc14} helps to maintain the blockchain of SolarCoin. The PoST allows users to stake their SolarCoin with a targeted yearly interest rate of 2\%. The PoST is a low energy algorithm in comparison to the Bitcoin’s PoW. When the PoST is set up on analogous scales, it spends less than 0.001\% of the power consumption compared to PoW.
	
		Solar energy is the focus of the SolarCoin. This is because solar energy does not produce surplus heat or carbon into the atmosphere. Moreover, the largest renewable energy source is solar energy. Other clean and renewable energy sources mostly need industrial implementations. In case of solar, solar panels can be used even by small groups or individuals to produce energy. With the technological advancements in this area, day-by-day the cost of solar energy is decreasing expeditiously.
		
		\subsubsection{NRGcoin}
		NRGcoin \cite{mihaylov2014nrgcoin, mihaylov2018nrgcoin} is actually an industry-academia project which was primitively developed at Vrije Universiteit Brussel. Presently, the Enervalis (www.enervalis.com) has scaled up this project in industrial content. NRGCoin helps to integrate green energy resources in the local grid by making it more beneficial for producers \& utilities and economical for both consumers \& government.
		
		NRGCoin differs from the other initiatives in these ways. In case of NRGCoin, the energy is not basically traded rather it is bought and sold by means of smart contract. Thus, this NRGCoin is not considered directly as a P2P energy trading approach. Moreover, NRGCoin does not only focus on solar energy and is not merely a cryptocurrency. Rather, it supports all kinds of renewable and clean resources.

		The three prominent components to the NRGCoin concept are the smart contract, currency market and gateway devices as illustrated in Fig. 11. The NRGCoin uses a novel blockchain-based smart contract which has replaced the traditional high-risk renewable supported policies. This smart contract runs on the Ethereum platform. Here, the consumers are supposed to offer 1 NRGCoin through the smart contract directly for every 1 kWh of produced green energy. On the other hand, from the coins paid by the consumers, the smart contract sends the grid fees as well as taxes to the DSOs.

		Various methods are used by the smart contract to substantiate the prosumer’s reported injection of green energy. New NRGCoins are forged by the smart contract once all the reports are checked out and audited. Moreover, prosumers are also rewarded for their injected green energy. It is totally up to the prosumers either they want to sell these coins on a currency market or use them afterward to pay for green energy. Consumers can buy their NRGCoins from the currency market to pay for their consumptions. The currency market enables prosumers to trade NRGCoins as other popular crypto- and non-cryptocurrencies such as the Euro, Bitcoin, and Dollar. This market can be set up as a new currency exchange platform, or it can be intermingled with the existing centralized or decentralized exchange markets.
		
		Though the mining rate decreases with time in order to prevent exorbitant inflation, the worth of NRGCoin remains the same as far as the amount of energy is traded. Here, 1 NRGCoin will always be 1 kWh produced energy no matter what the price of electricity will be. An NRGCoin that was bought five years back at low electricity price is the worth accurately the same amount of energy today. The NRGCoin can be used to pay for renewable and clean energies that are presently available in any local network.
		
		A primary advantage for prosumers is that an immutable blockchain assisted smart contract administers the disbursement of green energy which cannot be altered by market actors. For this reason, prosumers have blockchain-level pledges on their bounty for the energy they have contributed. The gateway devices along with the local electrical installation are set up in the local network, and they are connected to the Internet as well. The responsibilities of these gateway devices are to calibrate electricity inflows \& outflows, and also, connect with the blockchain and exchange market.

		\subsubsection{Electronic Energy Coin} Electronic Energy Coin (E2C) \cite{misc15} project is introduced as a green energy buying and selling platform which is developed by using blockchain and smart contract. This project strives for an energy revolution by ensuring a more secure, anonymous, fair, and proper energy distribution. E2C is built in accordance with the ERC-20 Ethereum token standard \cite{misc16}. Utilizing this standard makes E2C token faster in comparison to other cryptocurrencies. However, the smart contract provides a direct linkage between independent energy producers and consumers. Users can easily trade and exchange the E2C tokens for energy via the E2C platform. Moreover, E2C platform offers to predict the energy demand as well as supply dependant on the energy transactions. It also enables users to access the energy transactions so that they can make a much better and cognizant decision for future investments.

\begin{figure} [h]
	\includegraphics[width=\linewidth]{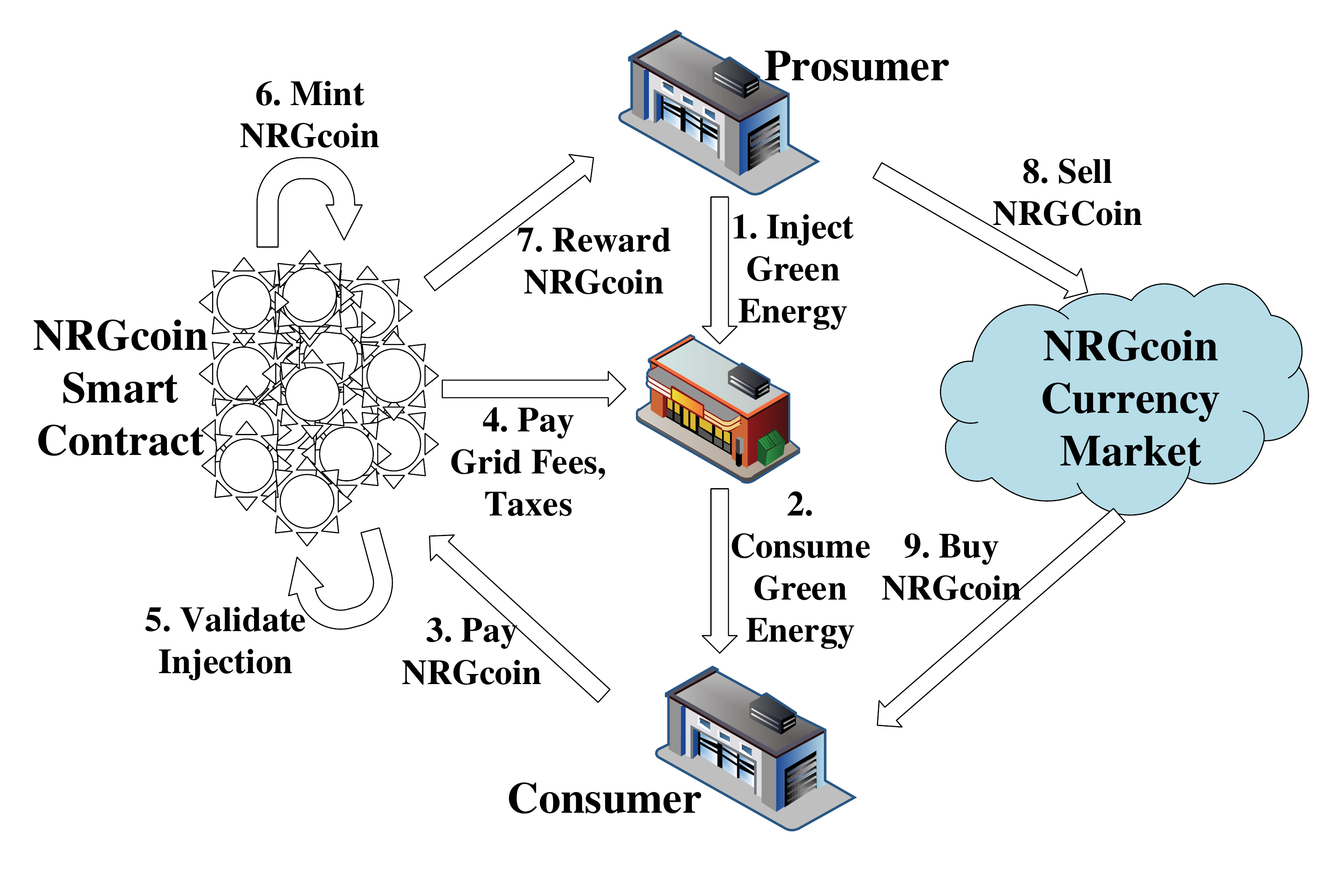}
	 \caption{The components of NRGcoin and their working procedures.}
\end{figure}

		\subsubsection{KWHCoin}
		KWHCoin \cite{misc17} is a blockchain-based cryptocurrency as well as a community which is supported by clean energy units. KWHCoin has a vision to lead the expansion of clean energy by reducing the cost of blockchain energy transactions. KWHCoin facts as an indigenous token for a decentralized application (DApp) where producers and consumers can link and set up their energy generation resources. In order to enact this, KWHCoin needs to build a platform that empowers people all over the world to buy and sell renewable energy resources with the help of “the Grid” which is a blockchain-based energy platform. In order to offer a comprehensive virtual company, the Grid utilizes a software and a collaboration among the peers. Once this virtual power grid is built, it will consistently perform without any hindrance and with no carbon footprint. Moreover, a 100\% clean energy will be delivered by the Grid directly from renewable and green energy resources. Users are free to select their energy providers as well as they can sell their generated clean energy to others with the help of the Grid in the form of KWH tokens.

		\subsubsection{TerraGreen Coin}
		TerraGreen coin \cite{misc18} is a unique blockchain-assisted initiative which manages biomass wastes from agriculture, home \& forestry sources and converts them into useful energy products. These products will have significant economic value once processed into energy. Generally, the purpose of this TerraGreen coin is to utilize the blockchain and cryptocurrency in order to make the earth a greener place, where the terra means the earth. 
		
		TerraGreen coin sets up on a consensus network and facilitates a fully decentralized P2P payment mechanism. However, in order to make the blockchain more energy efficient, the most recent dedicated PoS mechanism is utilized as a consensus mechanism instead of PoW. With lightning network, in TerraGreen blockchain, it takes only one second of block time to settle a new transaction and create a new block. Multiple data can be kept as well as tracked on the blockchain by means of a multilayered protocol. Moreover, users can follow the most recent project developments and can make their trading secure and easy via TerraGreen user interface (UI) and the secure wallet.
		
		One most important option of TerraGreen coin is that it supports unlimited sidechains in order to adopt other projects and cryptocurrencies in the umbrella of a green energy platform. TerraGreen coin enables users to participate directly in the management of biomass waste and also, in the production of renewable energy products, thus, contributing to the green energy revolution. Furthermore, the coins produced by the TerraGreen will be given away to the general public for the purpose of crowdfunding. The major biomass plant as well as the general public will be benefited from the investment made from this crowdfunding.

		\subsubsection{Charg Coin}
		Charg Coin \cite{misc19} is introduced by using blockchain in order to expedite the crowdsourced renewable energy distribution. This enables people to trade the energy each other within one-second. The Charg Coin blockchain facilitates the crowdsourcing of EV charging stations on the Internet of Energy (IoE) with the help of various partners of WeCharg (https://www.wecharg.com/). A common problem faced by all EVs is to find a charging station securely anytime and anywhere in any place. Charg Coin solves this problem with the help of WeChrag platform. WeCharg allows anyone at anyplace to join their EV charge station network so that everyone within the network can share their parking and charging stations with each other. By joining charge station owners with drivers, with this initiative, the environment pollution aware drivers do not need to worry about running out of power. Charg Coin can be traded in exchange of other popular currencies also. The charging stations on the network cannot be accessed or executed without having Charg Coin. Charg Coin is basically a tangible outcome of a smart document that runs on the Ethereum smart contract platform. Moreover, Charg Coin adopts a second layer blockchain solution as well to provide instant and micro-transactions at minimal cost.
		
		\subsubsection{CyClean Coin}
		CyClean coin \cite{misc20} is developed as a cryptocurrency with a vision to confront CO2 emission and promote clean energy by encouraging and provoking users with giving rewards. For CyClean coin, the Ethereum platform is employed to utilize smart contract. Moreover, CyClean coin adopts an innovative way of mining technique. They did not conform the conventional PoW mining technique to make the mining processing environmentally friendly since the conventional PoW mining spends an enormous amount of electricity to complete high computational tasks. Instead of PoW like mining, CyClean coin makes pre-mined all its coins in advance so that it can be ensured to provide rewards anytime to the users. Usually, only the users of CyClean products are guaranteed for this reward on a daily basis which includes electric motorbike, electric bicycle, electric car, and sunlight panel unit. For example, if anyone rides CyClean electric motorbike over a certain distance, that user will be awarded one CyClean coin.
	
	\subsection{Blockchain Platforms}
		\subsubsection{Pylon Network}
		Pylon Network \cite{misc21} is the exclusive and open-source platform which is designed and developed particularly for the needs of energy sector. The major objective of developing this technology is to accelerate the energy transition and to make sure that no one is left behind in the rapidly emerging era of prosumer participation, digitalization,  decentralization, and translucent cooperation in the area of energy. A neutral database based on blockchain technology is being developed in Pylon Network which allows to store and share energy data from the energy market stakeholders. With the help of Pylon Network database, users can determine with whom they want to share their energy data. Also, a new level of capability of competition can be achieved in the market with the help of data neutrality. The consumer, producer, and prosumer users can easily decide on their own private data. They also select the 3rd party service providers like ESCOs for them which are able to access their private information and provide the services back to them. This way can help the consumers to save on their bills as well. Fig. 12 represents a high-level model of Pylon Network blockchain.

		In Pylon Network, the data-sharing architecture authorizes the retailer companies and ESCOs who can grant cognizant and personalized services on the basis of high quality and granular consumer data. The blockchain technology of Pylon Network is offered an open-source platform in order to develop an inclusive communication mechanism that can be accessed by all stakeholders of the energy market. There are two assets in Pylon Network, one is the Pylon Token (PYLNT), and another one is the Pylon Coin (PYLNC). Both assets are related to each other and play an important role within the Pylon Network. Here, the Pylon Token can be defined as a digital asset in order to perform the engagement in the Pylon Network. On the other hand, the Pylon Coin is a digital currency that is awarded to the nodes for validating and creating the neutral database supported by the blockchain of Pylon Network. Simultaneously, some coins will go to the green energy projects to serve as a support for the people who are investing in the area of renewable technology. 
		
		Pylon Network’s blockchain discusses two crucial and pivotal views of blockchain technology. The first aspect is the energy waste minimization, whereas, the second is scalability. Instead of competitive mining, they have chosen cooperative mining which resulted in much lower consumption per transaction. The cooperative mining is being promoted against competitive mining for the purpose of improving energy efficiency and reducing the hardware cost as much as possible. Moreover, they are conforming to a series of tools and code lines that assure an impressive base of transactions per second. The on- and off-blockchain transactions could reach millions or even billions once the unification of all the tools is enforced. Furthermore, in case of instant payments, there is no need to be concerned about the block confirmation times as well as the transaction costs since the lightning network is utilized to transact and settle the off-chain.

\begin{figure} [h]
	\includegraphics[width=\linewidth]{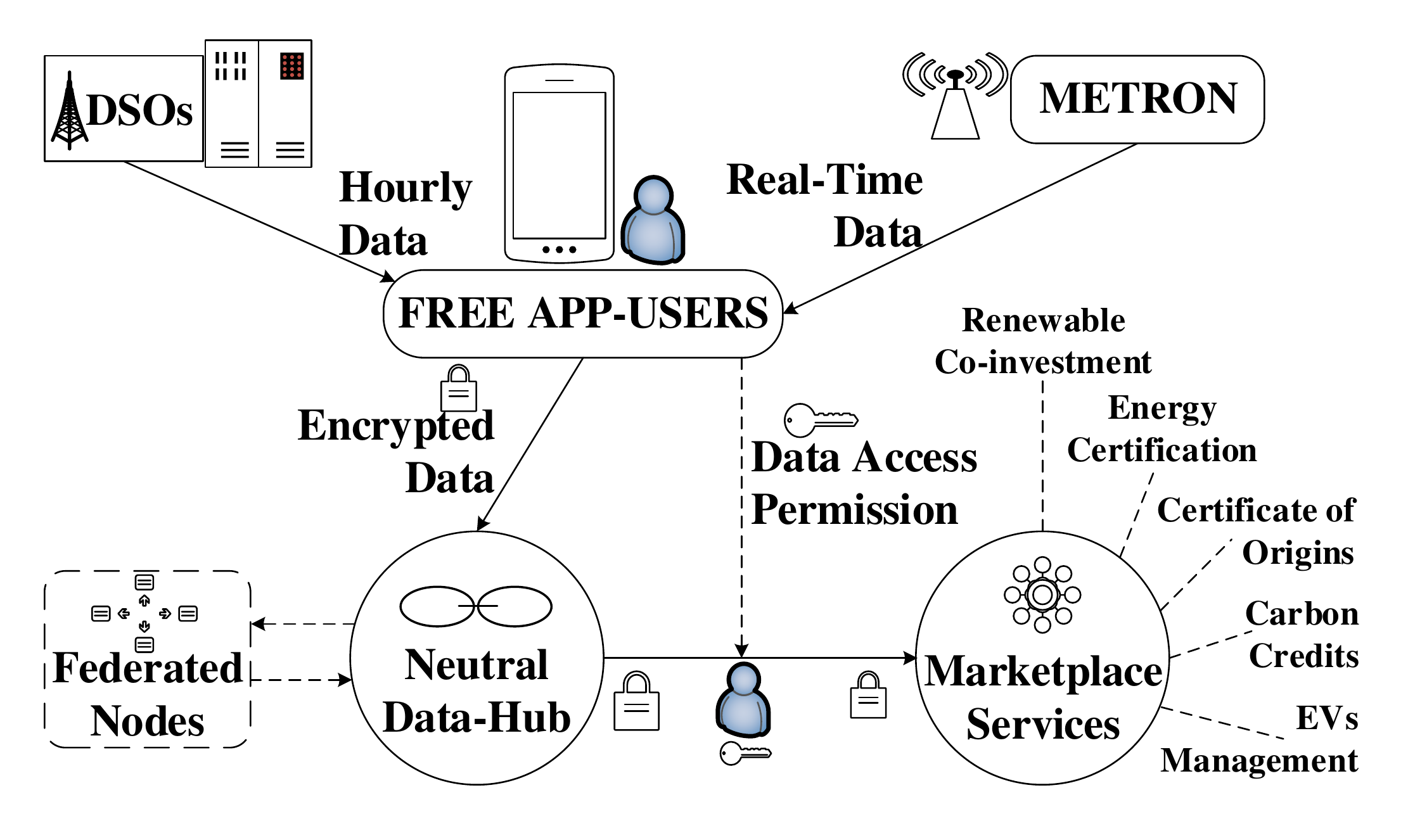}
	 \caption{A high-level model of the Pylon Network blockchain presented in \cite{misc21}.}
\end{figure}

		\subsubsection{EXERGY}
		The Exergy \cite{misc22} is developed by LO3 Energy Inc. (www.lo3energy.com) by means of blockchain and their own ingenious solutions. Exergy is a permissioned data platform for microgrids that constitutes local energy marketplace in order to transact energy across prevalent grid infrastructure. A mobile application is developed to set budget and receive alerts about energy availabilities. Through this Exergy platform, prosumers can trade their produced energy from their own renewable resources independently within their local marketplace. However, the role of the distributed system operator (DSO) is to regulate energy usage, load balancing, and demand response. Furthermore, in the case of EV charging, when there is an excess of energy in a public or private charging station or an EV, the excess energy can be traded on the local network.

		\subsubsection{Energy Web Chain} The Energy Web Foundation (www.energyweb.org/) is designed a public enterprise-grade blockchain platform and decentralized applications (DApps) for the energy sector. This platform is known as Energy Web Chain \cite{misc23}.
		
		Moreover, it is a public and permissioned Proof of Authority (PoA) blockchain which has high throughput in terms of processing the transactions. The Energy Web Chain is conversant with Ethereum developers as well as the open-source software development toolkit (SDK). These expedite the path to commercial decentralized applications.
		
		The Energy Web Chain introduces a first-layer utility token which is known as the Energy Web Token. This token is offered mainly for two major purposes. Firstly, it defends against any malicious activities. Secondly, it remunerates validators through transaction fees and block validation awards. The market participants can connect the assets from utility-scale generation to consumer-sided distributed energy resources (DER) through the Energy Web Chain. The assets with various hardware and software architectures are easily accommodated by the flexible SDK by means of full clients, light clients, and application programming interfaces (APIs). Thus, these components make Energy Web Chain easier for IoT device connection, grid balancing, EVs charging, and data authentication.

		\subsubsection{Powerledger}
		Powerledger \cite{misc24} is developed by utilizing blockchain in order to create a novel and transparent energy market in order to offer P2P renewable energy sharing. Within this Powerledger, the consumers and producers can set buy and sell prices to trade energies at appropriate and desired prices. The users can also keep the energies in battery storages so that it can be sold later time to get the maximum profit. Here, all energy transactions are stored in the blockchain in order to ensure more security. The trading of environmental commodities and renewable energy credits can also be possible through this Powerledger. The environmental commodities trading market is developing expeditiously. Consequently, there is more pressure to make sure that the credits are not double-counted or misused. Moreover, energy from renewable sources to offset emissions as well as the greater number of transactions related to environmental commodities and renewable energy credits can also be tracked.

		\subsubsection{Sunchain}
		Sunchain \cite{misc25} is developed to enable users to merge and share local solar energy by using consortium blockchain and IoT technologies. It is a solution for energy exchanges and meeting producers \& consumers. Like other initiatives, Sunchain is also going to contribute the transformation (fossil fuel to renewable) by providing blockchain-based solutions to developers and utilities in order to regulate energy exchanges. The blockchain architecture of Sunchain is developed for renewable energies, and also, it has extensive applications in smart grid. In this Sunchain, the blockchain keeps records of encrypted and signed data of smart meters. Moreover, it allows that the energy distribution among all the participants is actively handled and validated.

		Sunchain enforces energy sharing in accordance with the rules of the community. The ultimate vision of Sunchain is a sustainable development which is depicted by the nature of its consortium blockchain. Sunchain’s consortium blockchain does not use mining process, and hence, it consumes low electricity.
		
		Sunchain’s blockchain is tokenless, and it is not linked to any cryptocurrency. Also, it is designed to meet trust and scalability requirements. In addition, Sunchain’s blockchain pledges the origin of energy (eg., solar, other renewable sources). Because the energy amounts are written in unchangeable blockchain, this blockchain solution provides various services such as origin certification of renewable generations and traceability for energy consumptions.
		\subsubsection{ Dajie Blockchain Platform}
		Dajie blockchain platform \cite{misc26} is developed for microgrids which is based on IoT devices and blockchain technology. This platform allows microgrid community members to share their energies in peer-to-peer manner in their local neighborhood areas at a reasonable rate. A whole network of nodes is created by the IoT devices in a local micro-grid which enables users to exchange energy. However, to avail the energy coin and other facilities, users need install one of their IoT device and get themselves registered to their platform. Once registration is done, the Dajie platform will generate 1 energy coin for every kWh of energy produced. Finally, the energy coin generated will be stored in a secure wallet.

		Usually, small consumers and producers do not get reimbursed for their CO2 reduction contribution to the environment. This platform makes it possible for small consumers and prosumers to reclaim carbon credits through their introduced energy coin.

		\subsubsection{Greeneum Platform}
		Greeneum \cite{misc27} is a platform for renewable energy which is developed by utilizing machine learning, blockchain, smart contract, and IoT. The aim of this platform is to offer DApps, incentives for using renewable energies, and credits for reducing CO2 emissions. It deals with challenges of transforming from centrally grid-connected topology to regional community-based production and distribution via an integrated secure and decentralized solution.
		
		In order to record, manage, and trade renewable energies, Greeneum introduces Green tokens which are based on smart contracts. These tokens serve as the means of exchange and reward for the global community of green initiative supporters. With this platform, consumers can directly pay producers trough Green tokens. Moreover, Greeneum bonds are the additional tokens to award to the green energy producers. These producers may also get Greeneum carbon credits for utilization of green energy.
				
		Greeneum platform introduces two consensus mechanisms particularly for its energy applications such as Proof of Energy and Proof of Green. Besides these, in order to produce insights and precise predictions, machine learning algorithms are employed to help in grid optimization.

		\subsubsection{SunContract}
		SunContract \cite{misc28} is a blockchain-based platform which is implemented in Slovenia as a project. SunContract intends to maximize the benefits and prosperity of people, rather than utilities. Presently, Slovenian households are witnessing a reduction in their electricity costs using this platform. Moreover, they are able to select clean energy which opens a new business model for local energy trading without any monopoly.
		
		SunContract allows integrating different independent local energy producers and consumers. Via the SunContract mobile application, they can connect to the decentralized energy market directly. The SunContract platform is able to gain popularity among its users due to multiple reasons. With this platform, users can instantly access as well as audit the consumption and production insights. SunContract also ensures the transparency with the help of smart contract. Transparency of this platform helps to remove the need for intermediaries which are often used to create trust in the transactions and information transfer.
		\subsubsection{WePower}
		WePower \cite{misc29} is a platform for energy management and trading solutions. It introduces a blockchain-based token. It also offers various tools to facilitate users to know the energy usage pattern, look for a suitable renewable energy producer, make a contract with them digitally, and monitor the energy generations. Other than this, a financial mechanism named power purchase agreement, also called PPA, is introduced to enable direct energy transactions among renewable users. The main activities of WePower platform are presented in Fig. 13.
		
		With this solution, users can ensure stable costs based on their budgets. This platform also enables smaller companies having lower energy requirements to aggregate automatically with larger companies. In addition to this, these smaller companies can aggregate among themselves. Hence, this process creates a more extensive and comprehensive market. In such a market, opportunities and risks are shared among all, and thus, it can be managed in an efficient way. However, the token that introduced by WePower allows direct trade of PPA between the users.

\begin{figure} [h]
	\includegraphics[width=\linewidth]{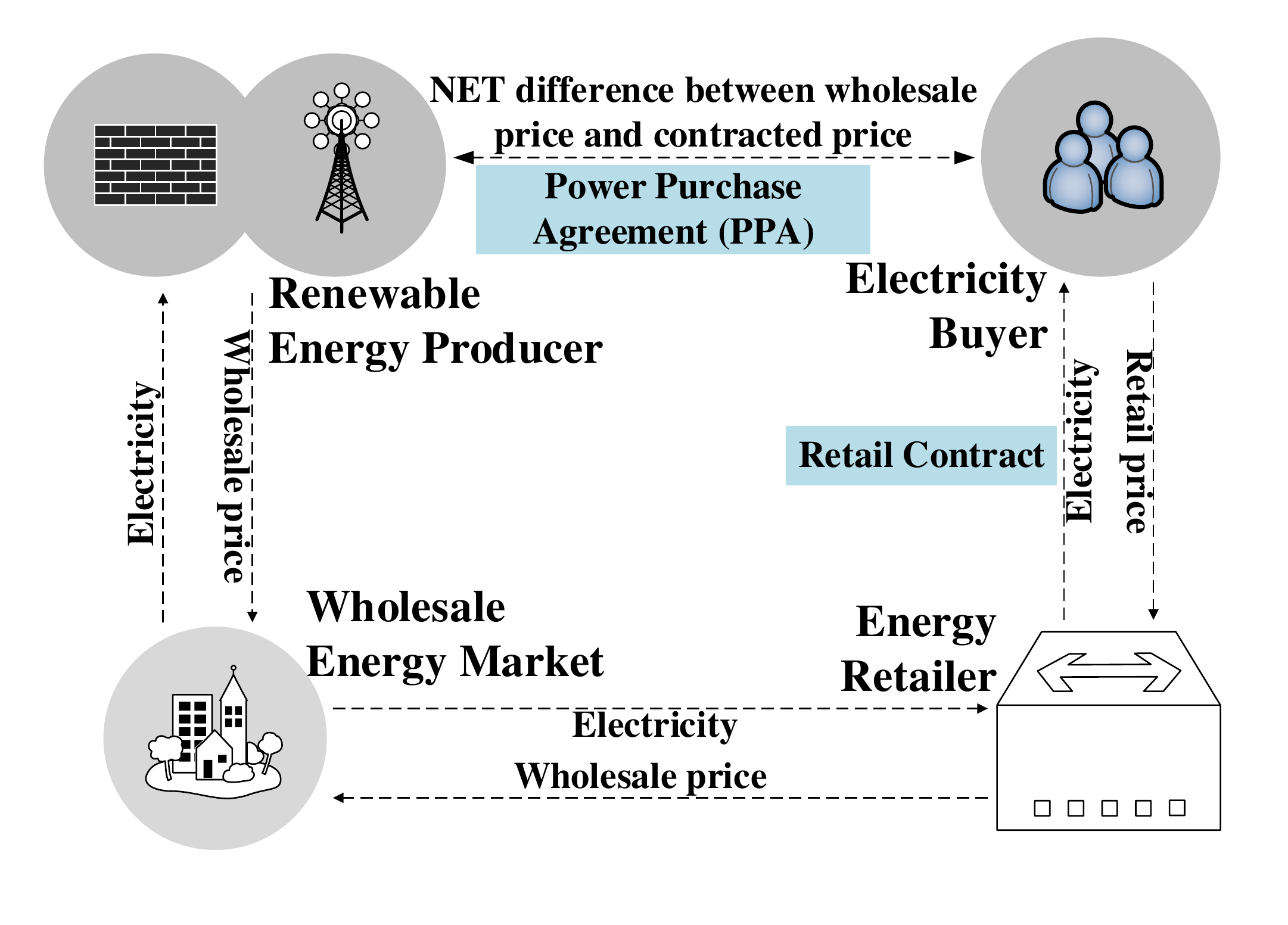}
	 \caption{The activities of WePower platform according to \cite{misc29}.}
\end{figure}

	\textit{Summary:} In this section, we have presented a number of recent blockchain initiatives for smart grid. We have divided this section into two sub-sections such as cryptocurrency initiatives and blockchain platforms. Next, we list a summary of these practical initiatives and cases in Table VI. From these recent initiatives, we have come to know that how these initiatives have adopted potential advantages of blockchain to provide decentralized services with security, privacy, and trust. However, among them the majority of works are focused on energy trading through public blockchain, though every initiative has its unique feature. On the other hand, keeping users’ information in public blockchain can lead towards privacy concerns. Hence, to ensure privacy of the energy producers and consumers should be in top priority. Moreover, with the rapidly growing blockchain-based smart grid initiatives, empowering users by taking control their own data should be consider seriously. In addition to these, the industries should work towards the inclusion of recent advancements of blockchain technology including sharding and off-chain like techniques to increase the transaction throughput substantially.
	
	\begin{table*}[ht]
\centering
\caption{Summary of Practical Projects and Trials}
\begin{tabular}{m{1.3cm}|m{.6cm}|m{2.5cm}|m{8.8cm}|m{2.8cm}}
\hline \hline
Name & Ref. & Organization & Description & Objective\\
\hline
SolarCoin & \cite{misc13} & SolarCoin Foundation & SolarCoin aims at providing rewards to solar power producers around the globe to reduce CO2 emission & Cryptocurrency for solar energy\\
\hline
NRGcoin & \cite{mihaylov2014nrgcoin, mihaylov2018nrgcoin} & Vrije Universiteit Brussel and Enervalis & NRGcoin is a decentralized virtual currency to use in renewable energy trading by prosumers in smart grid & Energy trading\\
\hline
Electronic Energy Coin (E2C) & \cite{misc15} & Electronic Energy & E2C is a blockchain-based secure, anynymous, and decentralized cryptocurrency for green energy trading which has an aim to enhance the control over energy transactions & Cryptocurrency for green energy\\
\hline
KWHCoin & \cite{misc17} & KWH Renewable Energy & KWHCoin is blockchain-based cryptocurrency for clean and renewable energy which converts physical units of kWh energy to digital tokens. Afterwards, people can purchase or sell energy through an another energy platform named The Grid in the form of digital tokens & Cryptocurrency for clean and renewable energy\\
\hline
TerraGreen Coin & \cite{misc18} & TerraGreen & TerraGreen Coin is developed particularly for biomass wastes and its converted form of energy products to support the clean energy revolution & Cryptocurrency for biomass energy\\
\hline
Charg Coin & \cite{misc19} & WeCharg & Charg Coin is desiged specifically for EVs that intends to offer crowdsources energy distribution among EVs through charing stations in the Internet of Energy (IoE) & Cryptocurrency for EVs\\
\hline
CyClean Coin & \cite{misc20} & CyClean Pte Ltd. & CyClean Coin aims at providing pre-mined coins as rewards to the users of CyClean products based on usages & Cryptocurrency for the CyClean products users\\
\hline
Pylon Network Blockchain & \cite{misc21} & Pylon Network & Pylon Network Blockchain is an energy blockchain platform where energy generations and usage data are stored, and the data owners have control to make decision who can access the data & Blockchain platform for energy sector/Energy sector\\
\hline
EXERGY & \cite{misc22} & LO3 Energy & EXERGY is a permissioned data platform which brings together energy and data to deploy localized marketplace for transating energy & Transactive energy\\
\hline
Energy Web Chain & \cite{misc23} & Energy Web Foundation (EWF) & Energy Web Chain is a decentralized and open-source blockchain platform developed for energy applications in order to accelerate low-carbon and user-centric energy transactions & Energy blokchcain platform\\
\hline
Powerledger & \cite{misc24} & Power Ledger Pty Ltd & Powerlerger is a blockchain-assisted platform to support emerging energy trading applications which offers transperancy while seamlessly interfacing with energy markets around the globe and also, interoperability between diverse pricing mechnisms and electricity units of different markets & Energy trading platform\\
\hline
Sunchain & \cite{misc25} & GreenTech Verte & Sunchain platform is designed for the management of solar enegy pooling and sharing by blockchain and IoT & Blokchain platform for solar energy exchange\\
\hline
DAJIE Blockchain Platform  & \cite{misc26} & Distributed Autonomous Joint Internet (DAJIE) Ltd. & DAJIE Blockchain Platform is designed for microgrid network where the community members (prosumers) are allowed to exchange energy in their local neighbourhood area at a better price than the grid. With this platform, prosumers are able to redeem to carbon credit as well & Blockchain platform for microgrid\\
\hline
Greeneum Platform & \cite{misc27} & Greeneum & Greeneum is a decentralized platform that faciliates DApps and incentive opportunities for contributing to reduce the CO2 emission by using renewable energy. Moreover, Greeneum provides accurate predictions to enable grid optimization using machine learning algorithms & Blockchain platform for renewable energy\\
\hline
SunContract & \cite{misc28} & SunContract & SunContract is an initiative implemented in Slovenia so that Slovanian households are able to buy and sell electricity freely, and this initiative has an aim to make those households energy self-sufficient & Peer-to-peer energy trading initiative\\
\hline
WePower Blockchain Platform & \cite{misc29} & WePower & WePower is a plarform for renewable energy contracting and trading that offers direct energy access, price certainty, cheaper transactions, and competitive rates to the energy users & Blockchain platform for renewable energy procurement and trading\\
\hline

\end{tabular}
\end{table*}

	\section{Research Challenges and Future Directions}
	We have identified in this paper that the blockchain presents many promising opportunities and applications for the future smart grid domain. The blockchain technology along with consensus mechanisms, smart contracts, and modern cryptographic techniques has made possible for entities to communicate without the support of any central authority or intermediary. However, blockchain-based systems consistently rely upon the veracity of the pre-determined rules despite the fact that no intermediaries are present during runtime and operation. Therefore, it is imperative to make sure that they are dependable, secure and precise. Furthermore, blockchain technology is yet at a growing stage and deals with various problems, such as diminished transaction loads.

	In addition to this, the intricacy of prevailing protocols and implementations is still challenging for researchers, users, and practitioners. On the other hand, though a number of works are being done, blockchain technology is still facing some other challenges as well which are caused by potential limitations due to its adoption in the smart grid.

	In this section, we present the implications of aforesaid research challenges to be addressed prior to its widespread implementation as well as adoption. We hope these six points will be worthy of future research directions towards integrating blockchain in the smart grid.

	\subsection{Blockchain Specific Challenges and Directions}

		\subsubsection{Throughput}
		Throughput in financial applications is defined as transaction processing time which is usually measured as the number transactions can be processed per second. Throughput in the blockchain is related to block interval time which remains a critical challenging issue in its implementation in cryptocurrency and also, other applications. This low throughput in blockchain opens a variety of challenges such as real-time transactions and micro-payment as well. Moreover, most successful and popular cryptocurrencies like bitcoin cannot be used directly in smart grid scenarios due to their low throughput. Notably, current public blockchains do not have enough high throughputs to compete with mainstream financial systems. Blockchain implemented as private and permissioned have a much higher throughput. However, these private blockchains are unable to provide total decentralization as they are usually deployed under a centralized control of any systems. Hence, it is desirable to access the tradeoff in-between decentralization and transactions speed for blockchain implementation.
		
		On the other hand, the off-chain technique is introduced to expedite the blockchain throughput, particularly for fast and micro-transactions. Though the off-chain technique is highly promising, it is still in its infancy, and further research efforts are necessary to do thorough examination and how to incorporate this new technique with smart grid. 
		
		Since in smart grid scenarios, the amount of data will probably be high, and both financial \& non-financial transactions will happen, addressing this throughput problem would be a big step forward in order to build the decentralized network for the smart grid.

		\subsubsection{Challenges with Consensus Mechanisms}
		As described previously in section III, the PoW is a computationally expensive consensus mechanism since it consumes a substantial amount of energy to confirm a transaction. Also, the PoS mechanism has a rich-rule problem. On the other hand, BFT-related mechanisms are not suitable for extensive public blockchain network where the number of participants is high. As a result, several consensus mechanisms have been developed to address the limitations and enhance the performance of currently popular mechanisms. Even though several research efforts have been made, we are certain that there is a leeway for more research, and the performance of newly proposed consensus mechanisms has not been rigorously analyzed. More research efforts should be dedicated to enhancing the performance that includes efficient, low energy consumption, high throughput, and highly scalable consensus mechanisms.
		
		The resource-constrained smart devices are not able to meet the substantial computational consumption to participate in consensus. Hence, the design of edge computing-assisted mechanisms will be one of the research challenges.
		
		One common limitation of all currently popular consensus mechanisms is their single purpose application like the usage in cryptocurrency only. While the benefits of blockchain will be common across all the non-cryptocurrency applications as well, the mechanisms should be different due to the nature of different applications. Thus, investigating the ways of blockchain consensus will continue to take part in future research for beyond cryptocurrency applications, particularly for different smart grid scenarios and, also, how to adopt currently popular mechanisms with the smart grid.

		\subsubsection{Blockchain and Smart Contract Security}
		Security is a never-ending game, and blockchain \& smart contract are no exception to this paradigm since both have some security problems. Particularly, public blockchain suffers more security problems than private blockchain. This is mainly due to engaging predefined entities in private blockchain. Moreover, smart contract may be vulnerable to malicious attacks, since the contracts are built on programming codes. These codes may contain vulnerabilities and bugs which open malicious parties to exploit. In \cite{saad2019exploring, li2020survey, conti2018survey} such malicious attacks against the blockchain and smart contract are described. Hence, it is crucial to design and implement the blockchain infrastructure and smart contracts which will be secure against attacks which could be future researches.
		
	\subsection{Smart Grid Centric Challenges and Directions}
	
		\subsubsection{Security and Privacy}
		In blockchain applications, security and privacy are the two considerable challenges that need to be addressed carefully. In the blockchain, the users usually are linked to public pseudonyms addresses to do anonymous transactions so that they can be unidentifiable and untraceable. To ensure transparency, all transaction-related information such as senders, receivers, records, amount of values are publicly accessible and available in blockchain. Consequently, though users use pseudonymous which may not help to stay completely anonymous them may lead towards security and privacy concerns.

		In smart grid scenarios, by analyzing open information, the users’ activities and energy profiles such as energy consumptions, productions, energy usage patterns, assets, and other records can be tracked and leaked. It can also reveal users’ real identities.
		
		Blockchain in smart grid may spread over large geographical areas. In the case of incorporating multiple chains and off-chains from different smart grid scenarios, the transactions and data exchange will occur through different chains. In this multichain platform, ensuring security and privacy will be complicated. Modern cryptographic schemes are capable of addressing these problems to limit access, exposure, and privacy. Already, many schemes have been developed to mitigate security and privacy concerns in blockchain.
		
		Traditional cryptographic security solutions usually built on centralized trusted entities that have a common scalability problem. On the other hand, blockchain applications deserve scalable and decentralized security solutions. As a result, many schemes are not directly suitable for blockchain applications. Adopting and designing cryptographic schemes particularly for blockchain in different smart grid scenarios while adapting with scenarios and ensuring the scalability along with efficiency need further research.

		\subsubsection{Incentive and Penalty Mechanisms}
		In a typical public blockchain network, once a miner or validator successfully generated a block, it usually receives a reward. In case of a group of mines or validators where they engage collaboratively, this reward is allocated among the participants. Unfortunately, lack of smart grid centric incentive mechanisms remains a challenging issue in their implementations. Particularly, providing incentives like any kind of rewards such as cryptocurrency, money, carbon credit, reputation value through blockchain ultimately helps generate, inject to the grid and consume in more clean \& renewable energies. Hence, a key future research direction is to design effective and strong incentive mechanisms with fair distribution to motivate all parties including producers, consumers and miner/validators to take part in blockchain network towards clean \& renewable energy usage. Moreover, penalty mechanisms are also essential in order to prevent malicious activities by any party.
		
		\subsubsection{Standardization}
		A number of mechanisms, protocols and technological solutions are being developed for the blockchain-based smart grid system. However, the primary challenge faced by the overall blockchain-based smart grid system is that it lacks widely accepted standards. Ultimately, this situation impedes the integration of smart meters, IoT devices, electric cyber-physical systems, EVs and other entities related to the smart grid. And, it also limits the interoperability among them. Hence, it is very essential to adopt the interoperability standards into the entire structure in order to make blockchain into smart grid reality. Moreover, to avoid disputes among different entities, new standardization efforts are essential since there is no trusted intermediary or a centralized authority like other systems. Therefore, the major objectives that can be achieved with blockchain standardization efforts in the context of smart grid system include communication exchange, advanced security \& privacy, reward/penalty policies, and seamless interoperability.

\section{Conclusion}
Blockchain technology in smart grid area is a new and emerging area of research that has attracted rapidly growing attentions. In this paper, we have presented a comprehensive studies of blockchain applications to future smart grid. Firstly, we have presented blockchain detailed background. Then, we have discussed about the future decentralized smart grid as well as blockchain features in order to understand the motivations of utilizing blockchain in smart grid security, privacy, and trust issues. Afterward, we have summarized recent blockchain contributions in smart grid. We have also summarized blockchain related practical initiatives for smart grid. Finally, we have outlined some challenges to direct future researches. We are hopeful that this paper will be a key step to open many possibilities for further research in this area.



\ifCLASSOPTIONcaptionsoff
  \newpage
\fi

\bibliographystyle{IEEEtran}
\bibliography{main}

\begin{IEEEbiography}[{\includegraphics[width=1in,height=1.25in,clip,keepaspectratio]{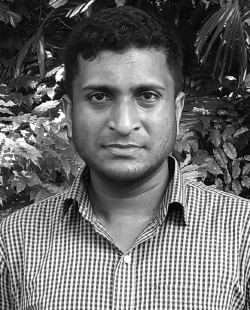}}]
{Muhammad Baqer Mollah} is currently working as a Research Associate in the Computer Science and Engineering at Nanyang Technological University (NTU), Singapore. Before joining NTU, he was working at Singapore University of Technology and Design (SUTD). He is currently involved in research works on AI and Blockchain applications to the cyber-physical systems (e.g., smart grid, industry, transportation). His research interests are mainly focused on advanced communication, security, and resource allocation techniques for future wireless networks and cyber-physical systems. He has a M.Sc. in Computer Science and B.Sc. in Electrical \& Electronic Engineering from Jahangirnagar University, Dhaka and International Islamic University Chittagong, Bangladesh, respectively.
\end{IEEEbiography}

\begin{IEEEbiography}[{\includegraphics[width=1in,height=1.25in,clip,keepaspectratio]{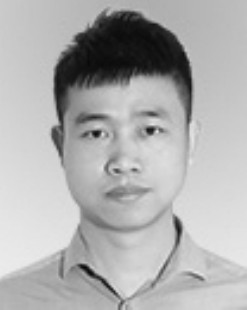}}]
{Jun Zhao} is an Assistant Professor in the School of Computer Science and Engineering at Nanyang Technological University (NTU). He received a PhD degree in Electrical and Computer Engineering from Carnegie Mellon University (CMU) in the USA and a Bachelor’s degree from Shanghai Jiao Tong University in China. Before joining NTU as a faculty member, he was a postdoctoral research fellow at NTU with Prof. Xiaokui Xiao. Prior to that, he was an Arizona Computing PostDoc Best Practices Fellow at Arizona State University working with Prof. Junshan Zhang therein and Prof. Vincent Poor at Princeton University. During the PhD training at CMU, he was advised by Prof. Virgil Gligor and Prof. Osman Yagan, while also collaborating with Prof. Adrian Perrig (now at ETH Zurich). His research interests include security/privacy (e.g., blockchains), wireless communications (eg., 5G, Beyond 5G/6G), and energy system (smart grid, Energy Internet, data center). He has over a dozen journal articles published in IEEE/ACM Transactions as well as over thirty conference/workshop papers.
\end{IEEEbiography}

\begin{IEEEbiography}[{\includegraphics[width=1in,height=1.25in,clip,keepaspectratio]{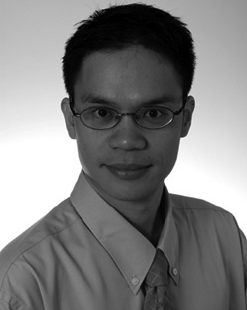}}]
{Dusit Niyato} is a Professor in the School of Computer Science and Engineering at Nanyang Technological University (NTU). He received B.Eng. from King Mongkut’s Institute of Technology Ladkrabang (KMITL), Thailand in 1999 and Ph.D. in Electrical and Computer Engineering from the University of Manitoba, Canada in 2008. He is a Fellow of IEEE. He received several best paper awards from well-known conferences such as IEEE ICC and IEEE WCNC. He is currently an editor of IEEE Transactions on Communications and IEEE Transactions on Vehicular Technology, and a senior editor of IEEE Wireless Communications Letter. His research interests are in the area of wireless communications and networks, game theory, smart grid, edge computing, blockchain technology, and Internet of Things (IoT).
\end{IEEEbiography}

\begin{IEEEbiography}[{\includegraphics[width=1in,height=1.25in,clip,keepaspectratio]{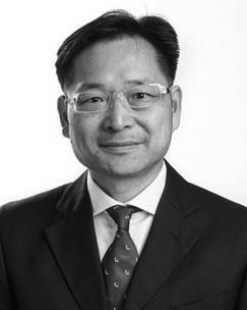}}]
{Kwok-Yan Lam} is currently a Professor at School of Computer Science and Engineering, Nanyang Technological University. Prior to joining NTU, he has been a Professor of the Tsinghua University, PR China (2002-2010) and a faculty member of the National University of Singapore and the University of London since 1990. He was a visiting scientist at the Isaac Newton Institute of the Cambridge University and a visiting professor at the European Institute for Systems Security. In 1998, he received the Singapore Foundation Award from the Japanese Chamber of Commerce and Industry in recognition of his R\&D achievement in Information Security in Singapore. He received his B.Sc. (First Class Hons.) in computer science from the University of London in 1987 and his Ph.D. from the University of Cambridge in 1990. His research interests include Distributed Systems, IoT Security Infrastructure, Distributed Protocols for Blockchain, Biometric Cryptography, Homeland Security, and Cybersecurity.
\end{IEEEbiography}

\begin{IEEEbiography}[{\includegraphics[width=1in,height=1.25in,clip,keepaspectratio]{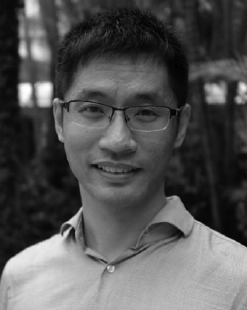}}]
{Xin Zhang} is currently an Assistant Professor in the School of Electrical and Electronic Engineering, Nanyang Technological University. He received the Ph.D. degree in electronic and electrical engineering from the Nanjing University of Aeronautics and Astronautics, China, in 2014, and the Ph.D. degree in automatic control and systems engineering from the University of Sheffield, U.K., in 2016. He was a Research Associate with the University of Sheffield, from 2014 to 2016, and the Postdoctoral Research Fellow with the City University of Hong Kong, in 2017. His research interests include power electronics, power system, and advanced control theory, together with their applications in various sectors. He was a recipient of the highly prestigious Chinese National Award for Outstanding Students Abroad, in 2016.
\end{IEEEbiography}

\begin{IEEEbiography}[{\includegraphics[width=1in,height=1.25in,clip,keepaspectratio]{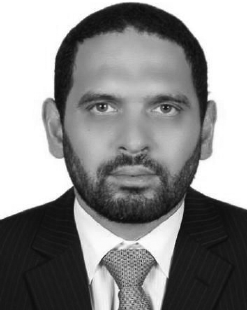}}]
{Amer M.Y.M. Ghias} is currently an Assistant Professor of the School of Electrical and Electronic Engineering, Nanyang Technological University, Singapore. He received the B.Sc. degree in electrical engineering from Saint Cloud State University, USA, in 2001, the M. Eng. degree in telecommunications from University of Limerick, Ireland, in 2006, and the Ph.D. degree in electrical engineering from the University of New South Wales (UNSW), Australia, in 2014. From February 2002 to July 2009, he had held various positions such as, Electrical Engineer, Project Engineer, and Project Manager, while working with the top companies in Kuwait. He has worked at UNSW, Australia (2014-2015) and the University of Sharjah, U.A.E (2015-2018). His research interests include hybrid energy storage, model predictive control of power electronics converter, fault-tolerant converter, modulations and voltage balancing techniques for multilevel converter, flexible AC transmissions and high voltage DC current.
\end{IEEEbiography}

\begin{IEEEbiography}[{\includegraphics[width=1in,height=1.25in,clip,keepaspectratio]{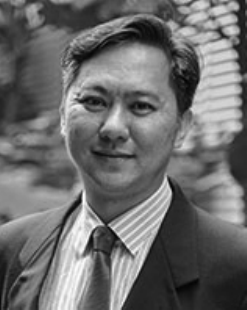}}]
{Koh Leong Hai} is currently a Senior Scientist at the Energy Research Institute @NTU (ERIAN), Nanyang Technological University, Singapore. He received the B.Eng. degree (Hons.) and the Ph.D. degree in electrical engineering from Nanyang Technological University (NTU), in 1994 and 2015, respectively. His current research interests include smart grid, energy information and management system, hybrid AC/DC microgrid, renewable energy and integration, and power system modeling and simulation.
\end{IEEEbiography}

\begin{IEEEbiography}[{\includegraphics[width=1in,height=1.25in,clip,keepaspectratio]{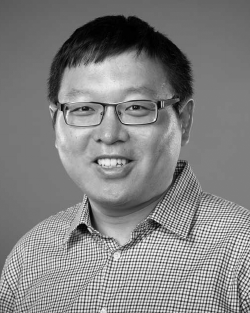}}]
{Lei Yang} is currently an Assistant Professor with the Department of Computer Science and Engineering, University of Nevada, Reno, NV, USA. He received the B.S. and M.S. degrees in electrical engineering from Southeast University, Nanjing, China, in 2005 and 2008, respectively, and the Ph.D. degree from the School of Electrical, Computer, and Energy Engineering, Arizona State University, Tempe, AZ, USA, in 2012, where he has been an Assistant Research Professor, since 2013. He was also a Postdoctoral Scholar with Princeton University, Princeton, NJ, USA. His research interests include stochastic optimization and big data analytics for renewable energy integration, grid integration of plugin electric vehicle, networked control of cyber-physical systems, modeling and control of power systems, network security and privacy, and network optimization and control. Dr. Yang received the Best Paper Award Runner-up of the IEEE INFOCOM 2014.
\end{IEEEbiography}

\end{document}